\documentclass[a4paper,fleqn,usenatbib]{mnras}

\usepackage[T1]{fontenc}
\usepackage{ae,aecompl}

\usepackage{graphicx}   
\usepackage{amsmath}    
\usepackage{amssymb}    

\title[A CEW model for SNe Ia]{A common-envelope wind model for Type Ia supernovae (I): binary
evolution and birth rate}

\author[Meng \& Podsiadlowski]{X. Meng$^{\rm 1, 2, 3}$ \thanks{E-mail: xiangcunmeng@ynao.ac.cn}\&
Ph. Podsiadlowski$^{\rm 4}$
\\
$^{1}$Yunnan Observatories, Chinese Academy of Sciences, 650216 Kunming, PR China\\
$^{2}$Key Laboratory for the Structure and Evolution of Celestial
Objects, Chinese Academy of Sciences, 650216 Kunming, PR China\\
$^{3}$Center for Astronomical Mega-Science, Chinese Academy of
Sciences, 20A Datun Road, Chaoyang District, Beijing, 100012, P.
R. China\\
$^4$ Department of Astronomy, Oxford University, Oxford OX1 3RH,
UK }

\date{Accepted XXX. Received YYY; in original form ZZZ}

\pubyear{2015}

\begin{document}
\label{firstpage}
\pagerange{\pageref{firstpage}--\pageref{lastpage}}
\maketitle

\begin{abstract}
The single-degenerate (SD) model is one of the principal models
for the progenitors of type Ia supernovae (SNe Ia), but some of
the predictions in the most widely studied version of the SD
model, i.e. the optically thick wind (OTW) model, have not been
confirmed by observations. Here, we propose a new version of the
SD model in which a common envelope (CE) is assumed to form when
the mass-transfer rate between a carbon-oxygen white dwarf (CO WD)
and its companion exceeds a critical accretion rate. The WD may
gradually increase its mass at the base of the CE. Due to the
large nuclear luminosity for stable hydrogen burning, the CE may
expand to giant dimensions and will lose mass from the surface of
the CE by a CE wind (CEW). Because of the low CE density, the
binary system will avoid a fast spiral-in phase and finally
re-emerge from the CE phase. Our model may share the virtues of
the OTW model but avoid some of its shortcomings. We performed
binary stellar evolution calculations for more than 1100 close WD
+ MS binaries. Compared with the OTW model, the parameter space
for SNe Ia from our CEW model extends to more massive companions
and less massive WDs. Correspondingly, the Galactic birth rate
from the CEW model is higher than that from the OTW model by
$\sim$30\%. Finally, we discuss the uncertainties of the CEW model
and the differences between our CEW model and the OTW model.

\end{abstract}




\section{Introduction}
Being an excellent cosmological distance indicators, Type Ia
supernovae (SNe Ia) have been successfully applied to determine
cosmological parameters (e.g. $\Omega$ and $\Lambda$;
\citealt{RIE98}; \citealt{PER99}). Indeed it has even been
proposed that SNe Ia can be used for testing the evolution of the
equation of the state of dark energy with time
(\citealt{HOWEL09}). SNe Ia also play an important part in
understanding the role of galactic chemical evolution as they are
believed to be the main producer of iron in their host galaxies
(\citealt{GM83}; \citealt{MATTEUCCI86}). They are also
accelerators of cosmic rays and sources of kinetic energy in
galaxy evolution processes (\citealt{HELDER09};
\citealt{POWELL11}).

It is widely accepted that SNe Ia originate from the thermonuclear
runaway of a carbon-oxygen white dwarf (CO WD) in a binary system.
The CO WD accretes material from its companion, increases mass to
its maximum stable mass, and then explodes as a thermonuclear
runaway. Almost half of the WD mass is converted into radioactive
$^{56}$Ni during the explosion (\citealt{BRA04}), and the amount
of $^{56}$Ni determines the maximum luminosity of SNe Ia
(\citealt{ARN82}). However, the precise nature of the progenitor
systems remains unclear (\citealt{BRANCH95}; \citealt{HN00};
\citealt{LEI00}; \citealt{MAOZ14}), although the identification of
the progenitor has wide-ranging implications in a number of
related astrophysical fields (\citealt{WANGB12};
\citealt{MENGXC15}). There have been two main competing scenarios
for the last 4 decades for the progenitor systems of SNe Ia, which
depend on the nature of the companion star. In the
single-degenerate (SD) model, a CO WD grows in mass by accretion
from its non-degenerate companion (\citealt{WI73};
\citealt{NTY84}) while, in the double degenerate (DD) scenario,
two WDs merge after losing angular momentum by gravitational-wave
radiation (\citealt{IT84}; \citealt{WEB84}).  There is some
support for both models, but there also are some serious problems,
both on the observational as well as the theoretical side
(\citealt{HOWEL11}; \citealt{MAOZ14}). In this paper, we focus on
the SD model.

\begin{figure*}
    \includegraphics[angle=90,scale=.65]{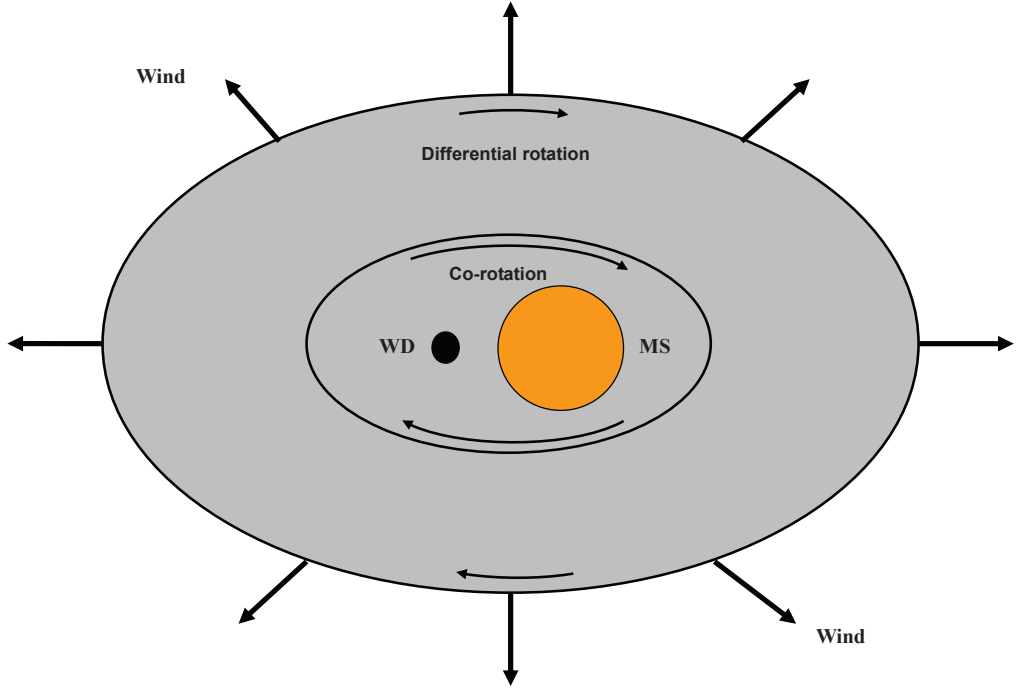}
    \caption{Schematic diagram illustrating the CE-wind model. The
mass-transfer rate from the Roche-lobe filling companion exceeds
the critical accretion rate at which the WD can accrete. This
leads to the formation of a common envelope (CE). The inner region
of the CE co-rotates with the binary, while the outer region
rotates differentially. Mass is lost from the surface of the CE,
powered in part by the luminosity emitted from the accreting WD
and in part by the frictional energy generated in the
differentially rotating region.}
    \label{ceshi}
\end{figure*}

In the standard SD model, the maximum stable mass of a CO WD is
close to the Chandrasekhar mass ($\sim 1.378\,M_{\odot}$,
\citealt{NTY84}). The companion of the WD may be a main sequence
star or a sub-giant star (WD+MS), or a red-giant star (WD+RG)
(\citealt{YUN95}; \citealt{LI97}; \citealt{NOM99};
\citealt{HAC99a, HAC99b}; \citealt{LAN00}; \citealt{HAN04}). There
is substantial observational support for the SD model: e.g., the
detection of circumstellar material (CSM) in the spectrum of SNe
Ia is usually taken as strong evidence in favor of the SD model
(\citealt{PAT07}; \citealt{STERNBERG11}; \citealt{DILDAY12};
\citealt{MAGUIRE13}). Moreover, supersoft X-ray sources (SSSs, WD
+ MS or WD + RG systems, \citealt{VANDERHEUVEL92};
\citealt{DIK03}) have been proposed to be good candidates as the
progenitors of SNe Ia (\citealt{HK03a, HK03b}). Recently, the UV
excesses expected from the collision between supernova ejecta and
their companion have been reported, which provides fairly
definitive support for the SD model at least in these cases
(\citealt{KASEN10}; \citealt{FOLEY12}; \citealt{BROWN14};
\citealt{CAOY15,CAOY16}; \citealt{MARION16}; \citealt{GRAUR16}).
It is therefore quite possible that all SN 2002es-like supernovae
arise from SD systems (\citealt{CAOY16}). A direct method for
confirming this type of progenitor model is to search for the
companion stars of SNe Ia in their remnants. The claimed discovery
of a potential companion of Tycho's supernova could in principle
support the WD + MS model (\citealt{RUI04}; \citealt{GONZA09};
\citealt{BEDIN14}), although there are some strong doubts about
this identification (\citealt{KERZENDORF09, KERZENDORF13}). Here,
we concentrate on the WD + MS channel, which is likely to be a
very important channel for producing SNe Ia in our Galaxy and
other late-type galaxies (\citealt{HAN04}).

Whether a CO WD explodes as a SN Ia in the SD model depends mainly
on the mass-transfer rate from its companion.  The maximum
accretion rate for stable hydrogen burning on a WD surface defines
a critical accretion rate, $\dot{M}_{\rm cr}$ (\citealt{NOMOTO82};
\citealt{NOM07}). If the mass-transfer rate is larger than
$\dot{M}_{\rm cr}$, the WD may expand to become a RG-like object,
and then a common envelope (CE) may form which engulfs both the WD
and its companion. In the past it had been suggested that the
system would then experience a spiral-in phase where the system
will ultimately merge and avoid a SN Ia (\citealt{NOMOTO79}); this
is one of the reasons for the low birth rate of SNe Ia from the SD
model. To avoid this problem, the optically thick wind (OTW) model
was proposed (\citealt{HAC96}; \citealt{HAC99a, HAC99b}), in which
the accreted hydrogen steadily burns on the surface of the WD at
the rate $\dot{M}_{\rm cr}$ while the unprocessed matter is lost
from the system as an OTW. Since \citet{HAC96}, many studies have
been based on the OTW model (\citealt{LI97}; \citealt{HAN04};
\citealt{CHENWC07}; \citealt{HKN08}; \citealt{MENG09};
\citealt{MENGYANG10}; \citealt{WANGB10}; \citealt{CHENXF11};
\citealt{HKSN12}). The OTW helps to modify the mass-growth rate of
the WD and the mass-transfer rate to avoid the formation of a CE,
increasing the birth rate of SNe Ia significantly
(\citealt{HAN04}). The OTW model may also explain the properties
of SSSs and recurrent novae (RN, \citealt{HK03a, HK03b};
\citealt{HK05, HK06a, HK06b}; \citealt{HKL07}).

However, some of the predictions of the model are in conflict with
observations. According to the opacity calculations by
\citet{IR96}, whether the OTW may occur or not is heavily
dependent on the Fe abundance. When $Z$ is lower than $0.002$, it
is found that the opacity is too low to drive a wind, i.e. no OTW
occurs (\citealt{KOB98}). This implies that there exists a
low-metallicity threshold for SNe Ia in contrast to SNe II.
However, no such metallicity threshold has been found in galaxy
host studies (\citealt{PRI07a}; \citealt{GALBANY16}). For the same
reason, SNe Ia are not expected at high redshift ($z>1.4$,
\citealt{KOB98}), but recently a SN Ia at $z=2.26$ was reported
(\citealt{RONDEY15}). Furthermore, some other SNe Ia at high
redshift and/or in low-metallicity environments have also been
discovered (\citealt{RONDEY12}; \citealt{FREDERIKSEN12};
\citealt{JONES13}). The material lost from the binary system in
the OTW will shape the CSM around the SN Ia; because of the large
wind velocity ($\sim1000\,{\rm km s^{\rm -1}}$), it may create a
low-density bubble or wind-blown cavity around the SN Ia.  Its
modification of the CSM on larger scales could become apparent
during the SNR phase. \citet{BADENES07} tried to search for the
signatures of wind-blown cavities in seven young SN Ia remnants
that would be expected in the OTW model. Unfortunately, they found
that such large cavities are incompatible with the dynamics of the
forward shock and the X-ray emission from the shocked ejecta in
all seven SN Ia remnants\footnote{We notice that
\citet{WILLIAMS11} reported results
  from a multi-wavelength analysis of the Galactic SN remnant RCW 86
  and found that the observed characteristics of RCW 86 may be
  reproduced by an off-center SN explosion in a low-density cavity
  carved out by the progenitor system. However, the remnant has
  usually been considered a core-collapse SN
  (e.g. \citealt{GHAVAMIAN01}) although this is still being debated
  (\citealt{BROERSEN14}).}.  This result may indicate that the
progenitors do not modify their surroundings in a strong way --
especially, the absorption features seen in the spectra of SNe Ia
suggest small cavities of radius $\sim10^{\rm 17}$ cm
(\citealt{PAT07}; \citealt{BLONDIN09}; \citealt{BORKOWSKI09};
\citealt{SIMON09}; \citealt{STERNBERG11}; \citealt{PATNAUDE12}).
In addition, by modelling two remnants in the Large Magellanic
Cloud with strong Fe-L line emission in their interiors,
\citet{BORKOWSKI06} found that the two remnants required a large
interior density, which would be expected from a low-velocity
pre-explosion wind rather than the fast OTW.

As an alternative, a super-Eddington-wind (SEW) scenario has been
suggested which only weakly depends on metallicity
(\citealt{MAX13}). However, this model also has the wind velocity
problem. In addition, whatever in SEW model or in OTW model, there
is a fine-tuning problem that how the helium flash and hydrogen
burning adjust with each other (e.g. \citealt{PIERSANTI00};
\citealt{SHEN07}; \citealt{WOOSLEY11}). This issue is crucial to
determine whether or not a CO WD may effectively increase its
mass.

\begin{figure*}
    \includegraphics[angle=270,scale=.65]{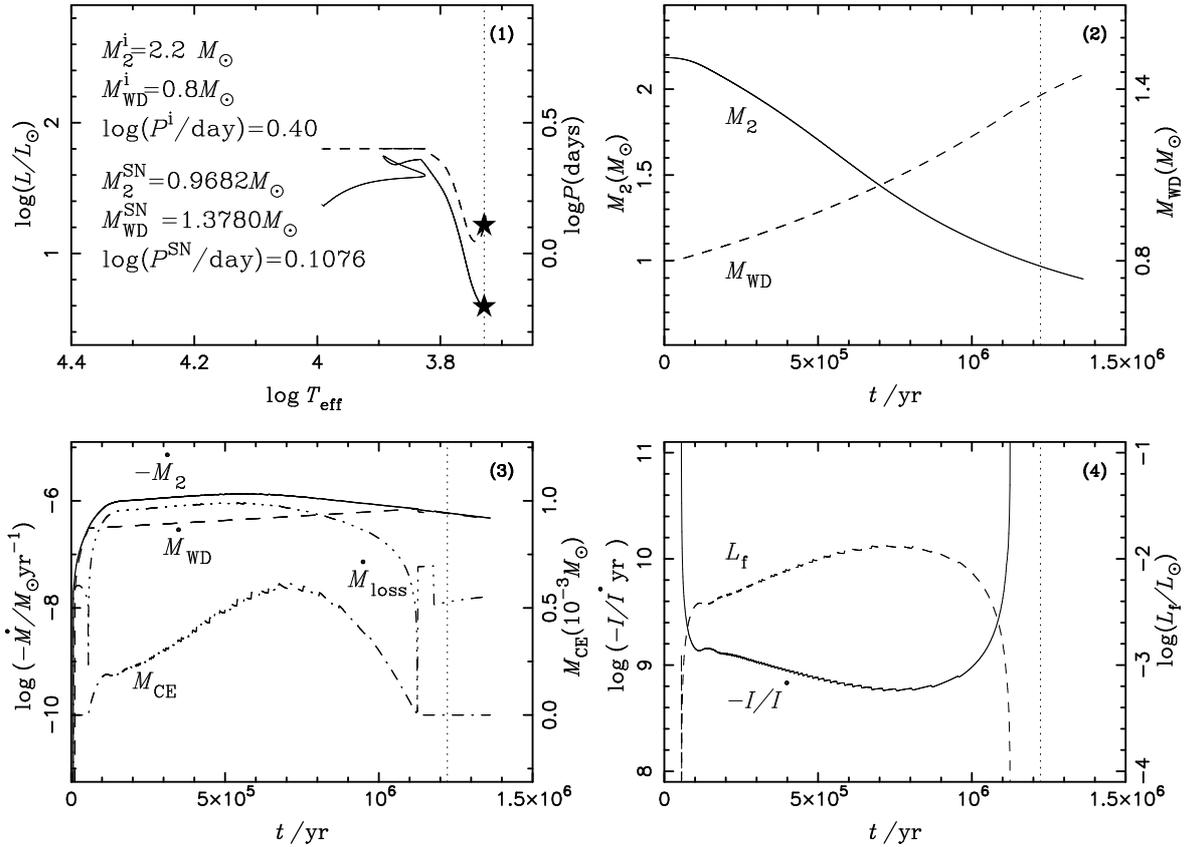}
    \caption{An example of the evolution of a binary in the CEW model.
The evolution of various parameters are shown, including the CO WD
mass, $M_{\rm WD}$, the secondary mass, $M_{\rm 2}$, the
mass-transfer rate, $\dot{M}_{\rm 2}$, the mass-growth rate of the
CO WD, $\dot{M}_{\rm WD}$, the mass of the CE, $M_{\rm CE}$, the
mass-loss rate from the system, $\dot{M}_{\rm loss}$, the
frictional luminosity, $L_{\rm f}$, and the merger timescale for
the binary system, $I/\dot{I}$, as labelled in each panel. The
evolutionary tracks of the donor stars are shown as solid curves
and the evolution of the orbital periods is shown as dashed curves
in panel (1). Dotted vertical lines in all panels and asterisks in
panel (1) indicate the position where the WD is expected to
explode as a SN Ia. The initial binary parameters and the binary
parameters at the time of the SN Ia explosion are also given in
panel (1).}
    \label{ce0822}
\end{figure*}

If there is no OTW or SEW, the status of the SD model goes back to
the situation some 20 years ago. However, whether the formation of
a CE really needs to be avoided, as suggested by \citet{NOMOTO79},
has never been properly addressed. In this paper, we constructed a
new SD model in which no OTW occurs but a CE forms around the
binary systems when the mass-transfer rate exceeds the critical
accretion rate. As we will show, a CE does not generally lead to
the merger of the binary system as the density in the CE is
usually quite low, resulting in a long spiral-in timescale. Indeed
the existence of a temporary CE has many attractive features: the
WD will naturally grow at the stable nuclear burning rate, both
for H and He shell burning phases (similar to the situation in
thermally pulsing asymptotic giant branch [TPAGB] stars), avoiding
some of the fine-tuning in the classical SD model. While the CE
acts as a mass reservoir, the wind from an extended CE envelope
will naturally produce the low velocities in the circumstellar
medium (CSM) as inferred for some SNe Ia (\citealt{BADENES07};
\citealt{PAT07}). As the system will have the appearance of a
TPAGB star in the main WD accretion phase, no X-rays are expected,
alleviating the existing X-ray constraints on the SD model
(\citealt{GB10}; \citealt{DISTEFANO10}). In many cases, some
H-rich CE material is still left at the time of the explosion.
While, in most cases, this will not be directly detectible, it may
explain the high-velocity Ca features observed in many SNe Ia
which seem to require the existence of some H
(\citealt{MAZZALI05a,MAZZALI05b}).

In this paper we will comprehensibly study this new framework for
the SD scenario specifically for the WD + MS channel and
systematically determine the parameter space for potential SN Ia
progenitors.  The results can be applied to study the statistical
properties of SNe Ia using a binary population synthesis approach
and may be helpful in searches for potential progenitor systems of
SNe Ia. In section \ref{sect:2}, we describe the detailed
numerical method for the binary evolution calculations and the
model grid we have calculated. The results of these calculations
are presented in section \ref{sect:3}. Our binary population
synthesis (BPS) is presented in section \ref{sect:4}, and the BPS
results are shown in section \ref{sect:5}. We briefly discuss our
results in section \ref{sect:6}. Finally, we summarize the main
results in section \ref{sect:7}, where we also discuss the future
work required to develop the model further.

\section{The common-envelope wind model for SNe Ia}\label{sect:2}
\subsection{Physics input}\label{subs:2.1}
To calculate the binary evolution of WD+MS systems in detail, we
adopt the stellar evolution code developed by \citet{EGG71, EGG72,
EGG73}. During the last four decades, the code has been updated
repeatedly with the latest input physics (\citealt{HAN94};
\citealt{POL95, POL98}). For example, Roche-lobe overflow (RLOF)
is treated as a modification of one boundary condition to ensure
that the companion overfills its Roche lobe but never much
overfills its Roche lobe for the steady RLOF (\citet{HAN00}). The
ratio of mixing length to local pressure scale height,
$\alpha=l/H_{\rm p}$, is set to be 2.0, and the convective
overshooting parameter, $\delta_{\rm OV}$, to be 0.12
(\citealt{POL97}; \citealt{SCH97}), which roughly equals to an
overshooting length of $0.25 H_{\rm P}$. Solar metallicity, i.e.
$Z=0.02$, is adopted in this paper. The opacity tables have been
compiled by \citet{CHE07} from \citet{IR96} and \cite{AF94}.

\begin{figure}
    \includegraphics[angle=270,scale=.65]{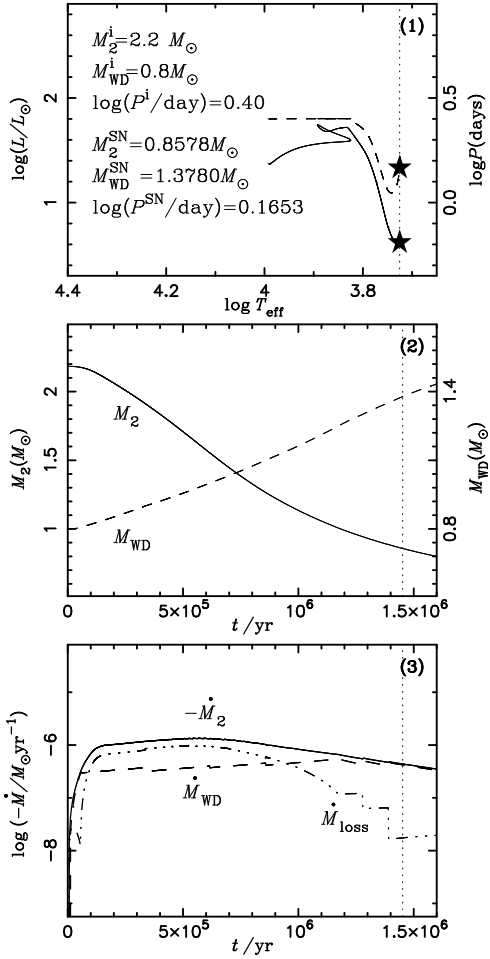}
    \caption{An example of a binary evolution sequence in the OTW
model. The evolution of various parameters are shown here,
including the CO WD mass, $M_{\rm WD}$, the secondary mass,
$M_{\rm 2}$, the mass-transfer rate, $\dot{M}_{\rm 2}$, the
mass-loss rate from the system, $\dot{M}_{\rm loss}$, and the
mass-growth rate of the CO WD, $\dot{M}_{\rm WD}$ as labelled in
each panel. The evolutionary tracks of the donor stars are shown
as solid curves, and the evolution of the orbital period is shown
as a dashed curve in panel (1). Dotted vertical lines in all
panels and asterisks in panel (1) indicate the position where the
WD is expected to explode as a SN Ia. The initial binary
parameters and the binary parameters at the time of the SN Ia
explosion are also given in panel (1).}
    \label{ha0822}
\end{figure}

\subsection{Model}\label{subs:2.2}
In the WD + MS channel\footnote{We also plan to apply the CEW
model to the WD + RG channel in the future, once the model has
been developed sufficiently, as the WD + RG channel may be
essential for producing SNe Ia in old populations. See section
\ref{subs:6.1} where we discuss some of the key uncertainties in
our model.}, the companion fills its Roche lobe on the main
sequence or in the Hertzsprung gap (HG) and transfers material
onto the WD. If the mass-transfer rate, $|\dot{M}_{\rm 2}|$,
exceeds the critical value, $\dot{M}_{\rm cr}$, the WD will become
a RG-like object and fill and ultimately overfill its Roche lobe.
A common envelope (CE) then forms. Hereafter, we shall refer to
our model as the CE wind (CEW) model to distinguish it from the
original OTW model (Fig.~\ref{ceshi} provides a schematic cartoon
of the CEW model)\footnote{Note that our model differs
significantly from the core-degenerate model, in which the CE
forms on a dynamical timescale and the newly formed binary in the
CE consists of the WD and the hot core of an AGB star
(\citealt{SOKER11}; \citealt{ILKOV12,ILKOV13}), while for our CEW
model, the CE is maintained on a thermal timescale, and the binary
embedded in the CE consists of a WD + MS system.}.
For the CE structure, our model is similar to that in
\citet{MEYER79}. In the model of \citet{MEYER79}, the CE is
divided into two regions, i.e. a rigid rotating inner region and a
differentially rotating outer region with a simply assumed sharp
boundary at distance $s\simeq a$, where $a$ is the binary
separation. The inner region co-rotates with the binary, while the
angular frequency of the outer region drops off as a power law,
which means that at the boundary between the inner and outer
regions, the angular frequency of the CE equals the orbital
angular frequency of the inner binary system. Here, we also
assume, as in \cite{MEYER79}, that the inner region of the CE
co-rotates with the binary while the outer region rotates
differentially. The differentially rotating outer envelope
continually extracts the orbital angular momentum from the inner
binary to the slowly rotating outer part at a rate
 \begin{equation}
 \dot{I}=-\alpha\eta a^{\rm 3}\Omega,\label{eq:dotang}
  \end{equation}
where $a$ is the binary separation, $\Omega$ is the Keplerian
orbital angular frequency of the WD + MS binary,

 \begin{equation}
 \Omega^{\rm 2}=\frac{G(M_{\rm 1}+M_{\rm 2})}{a^{\rm 3}},
  \end{equation}
$\eta$ is the effective turbulent viscosity,
 \begin{equation}
 \eta=\rho lv_{\rm c}/2\label{eq:eta}
  \end{equation}
($\rho$ is the CE density, $v_{\rm c}$ the convective velocity,
$l$
 the mixing length) taken at the boundary of the co-rotating inner
 region, i.e. $s\simeq a$, and $\alpha$ is a coefficient of order one. Then, the
 friction between the inner and the outer regions of the CE leads to a
 decrease of the orbital angular momentum
 \begin{equation}
 I=\frac{M_{\rm 1}M_{\rm 2}}{M_{\rm 1}+M_{\rm 2}}a^{\rm 2}\Omega,\label{eq:ang}
  \end{equation}
and the energy released by the shrinking binary orbit  is
 \begin{equation}
 L_{\rm f}=\frac{d}{dt}\left(\frac{GM_{\rm 1}M_{\rm
 2}}{2a}\right)=\alpha\eta G(M_{\rm 1}+M_{\rm 2}).\label{eq:lf}
  \end{equation}
Actually, almost all of the potential energy released is
dissipated as frictional heat and added to the nuclear energy to
expand the CE, where $\alpha$ represents the efficiency of
transferring the orbital angular momentum of the inner binary
system to the angular momentum of the outer CE (See
equation~\ref{eq:dotang}). Here, guided by the results of
\citet{MEYER79}, we\footnote{At present, the CE model is still
very simple and many parameters are quite uncertain, especially
the frictional density. Here, the chosen values for some of the
parameters are just set for guidance to mainly set the scale of
the frictional density. Generally, $\eta\propto l$; the mixing
length will be much less than $a$ but a significant fraction of it
and we, somewhat arbitrarily, set $l=0.1\,a$. Our choice of values
for $\alpha$ and $l$ results in $\alpha\eta\propto a$, which is of
the same order as estimated by \citet{MEYER79} ($\alpha=6\pi$ in
\citealt{MEYER79}). Therefore, the uncertainties of $\alpha$ and
$\eta$ are transferred into an uncertainty in the frictional
density. We will discuss the influence of the frictional density
on our model in detail in section \ref{subs:6.1}.} simply set
$\alpha=10$, $l=0.1\,a$ and $v_{\rm
  c}=0.26\sqrt{GM_{\odot}/a}$. We also treat the CE as spherical and
$\rho$ in equation (3) is set to be the average density of the CE
 \begin{equation}
\overline{\rho}=\frac{M_{\rm CE}}{\frac{4}{3}\pi R_{\rm CE}^{\rm
3}},\label{eq: rho}
  \end{equation}
where $M_{\rm CE}$ is the CE mass, which is derived from
 \begin{equation}
\dot{M}_{\rm CE}=|\dot{M}_{\rm 2}|-\dot{M}_{\rm WD}-\dot{M}_{\rm
wind}.\label{eq:mcedot}
  \end{equation}
$\dot{M}_{\rm wind}$ is the mass-loss rate from the CE surface due
to a wind, which is obtained by modifying the Reimer's wind
formula (\citealt{REIMERS75}):
 \begin{equation}
\dot{M}_{\rm wind}=1\times10^{\rm -13}\frac{(L_{\rm
tot}/L_{\odot})(R_{\rm CE}/R_{\odot})}{(M_{\rm 1}+M_{\rm 2}+M_{\rm
CE})/M_{\odot}}\, M_{\odot}\,{\rm yr}^{\rm -1},\label{eq:wind1}
  \end{equation}
where $L_{\rm tot}=L_{\rm nuc}+L_{\rm 2}+L_{\rm f}$. Here, $L_{\rm
2}$ is the secondary luminosity and $L_{\rm nuc}$ is the nuclear
energy for stable hydrogen burning
 \begin{equation}
 L_{\rm nuc}=0.007X\dot{M}_{\rm cr}c^{\rm 2},\label{eq:Lnu}
  \end{equation}
where $X$ is the hydrogen mass fraction.

\begin{figure*}
\vspace{5mm}
    \includegraphics[angle=270,scale=.65]{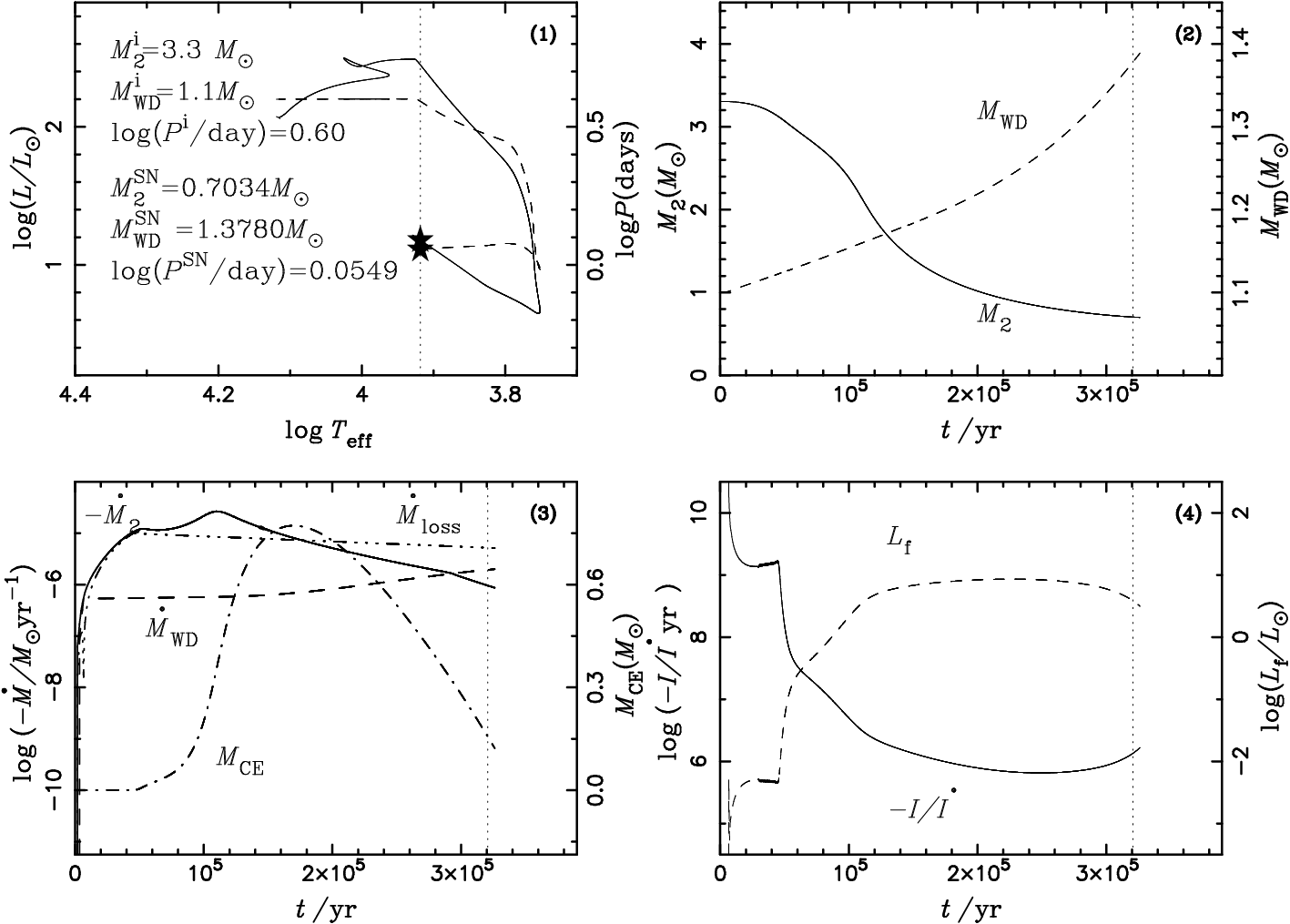}
\caption{Similar to Fig.~\ref{ce0822} for different initial binary
parameters.} \label{110633}
\end{figure*}

Assuming that the system can be modelled as a red-giant star, the
effective temperature, radius and mass of system approximately
follow the relation
 \begin{equation}
 \frac{T_{\rm eff}}{T_{\rm eff, \odot}}=\left(\frac{R_{\rm CE}}{R_{\odot}}\right)^{\rm
 -0.1}\left(\frac{M_{\rm WD}+M_{\rm 2}+M_{\rm CE}}{M_{\odot}}\right)^{\rm 0.1},\label{eq:rg}
  \end{equation}
where $T_{\rm eff, \odot}$ and $R_{\odot}$ are the solar effective
temperature and radius, respectively (\citealt{WUTAO14}). The
effective temperature of the CE should be between 2500\,K and
3200\,K (\citealt{SCHRODER07}). For simplicity, we set the
effective temperature of the CE to 3000\,K; then the radius of the
CE\footnote{The radius of the CE in the CEW model may be larger
than 500\,$R_{\odot}$; then the wind velocity from the CE surface
is very likely to be lower than 50 km/s, consistent with the
observations of the variable Na absorption lines in the spectrum
of some SNe Ia (\citealt{PAT07}; \citealt{SIMON09}).} can be
obtained from equation (\ref{eq:rg}). Based on Equations
(\ref{eq:mcedot}), (\ref{eq:wind1}) and (\ref{eq:rg}), we may
calculate the CE mass as

 \begin{equation}
 M_{\rm CE,i+1}=M_{\rm CE, i}+\dot{M}_{\rm CE,i}\cdot\Delta t,\label{eq:cemass}
  \end{equation}
where $\Delta t$ is the time step in the binary evolution
calculation, and $M_{\rm CE, 0}=0$ before the mass-transfer rate
exceeds the critical accretion rate. (See the Appendix for details
on how the CE mass is calculated in practice.)

$\dot{M}_{\rm WD}$ in equation (\ref{eq:mcedot}) is the
mass-growth rate of the WD, which depends on the critical
accretion rate (\citealt{HAC99a})
 \begin{equation}
 \dot{M}_{\rm cr}=5.3\times 10^{\rm -7}\frac{(1.7-X)}{X}(M_{\rm
 WD}-0.4)\, M_{\odot}\,{\rm yr}^{\rm -1},\label{eq:mdotcr}
  \end{equation}
where $X$ is the hydrogen mass fraction and $M_{\rm WD}$ the mass
of the accreting WD (in $M_{\odot}$). If the CE exists, the
structure of the WD is similar to a TPAGB star, and we therefore
set $\dot{M}_{\rm WD}=\dot{M}_{\rm cr}$, where hydrogen is stably
burning into helium, and helium flash and hydrogen burning phases
adjust themselves just as they do in a TPAGB star. When the
mass-loss rate from equation (\ref{eq:wind1}) is very high, a CE
cannot be maintained even if the mass-transfer rate is higher than
the critical accretion rate. In this case, we set $\dot{M}_{\rm
wind}=|\dot{M_{\rm 2}}|-\dot{M}_{\rm WD}=|\dot{M_{\rm
2}}|-\dot{M}_{\rm cr}$. In the OTW model, when $|\dot{M}_{\rm
2}|>\dot{M}_{\rm cr}$, $\dot{M}_{\rm WD}=\eta_{\rm He}\dot{M}_{\rm
cr}$, where $\eta_{\rm He}$ is the mass accumulation efficiency
for helium flashes. Actually, even if a WD accretes hydrogen-rich
material at the rate of $\dot{M}_{\rm cr}$, helium burning is
always unstable, i.e. helium flashes occur, leading to $\eta_{\rm
He}<1$ even though $|\dot{M}_{\rm 2}|>\dot{M}_{\rm
cr}$(\citealt{KH2004}). In section \ref{subs:6.2}, we will discuss
the effect of $\eta_{\rm He}$ on the results of our model. So, in
our CEW model, the mass-growth rate of the WD during the phase
when $|\dot{M}_{\rm 2}|>\dot{M}_{\rm cr}$ is higher than that in
the classical OTW model.

The total luminosity may exceed the Eddington luminosity of the
system,
 \begin{equation}
 \displaystyle
  \begin{array}{lc}
 L_{\rm Edd}=\frac{\displaystyle 4\pi G \left(M_{\rm WD}+M_{\rm 2}+M_{\rm CE}\right)c}{\displaystyle \kappa_{\rm T}}\\
\\
 \hspace{0.75cm} \simeq3.3\times10^{\rm 4}\frac{\displaystyle M_{\rm WD}+M_{\rm 2}+M_{\rm CE}}{\displaystyle M_{\odot}}\, L_{\odot},
 \end{array}
  \end{equation}
where $c$ is the speed of light and $\kappa_{\rm T}$ is the
Thomson opacity. When the total luminosity exceeds the Eddington
luminosity, the common envelope will expand and lose its material
at a high mass-loss rate. Both of these effects decrease the CE
density and hence the frictional density between the binary system
and the CE. As a consequence, the frictional luminosity will
decrease until $L_{\rm tot}=L_{\rm Edd}$. However, the mass-loss
rate and/or the expanding speed of the CE are quite uncertain, and
it is difficult to obtain the frictional density ($L_{\rm f}$)
directly. Since the CE may self-regulate to maintain $L_{\rm
tot}\simeq L_{\rm Edd}$ by expanding or losing material at a high
mass-loss rate, we set the frictional luminosity to be $L_{\rm
f}=L_{\rm Edd}-L_{\rm nuc}-L_{\rm 2}$ if the total luminosity
exceeds the Eddington luminosity of the system, and
 \begin{equation}
\frac{\dot{I}}{I}=-\frac{L_{\rm f}a}{GM_{\rm WD}M_{\rm 2}},
  \end{equation}
and
 \begin{equation}
\dot{M}_{\rm wind}=1\times10^{\rm -13}\frac{\displaystyle (L_{\rm
Edd}/L_{\odot})(R_{\rm CE}/R_{\odot})}{\displaystyle (M_{\rm
1}+M_{\rm 2}+M_{\rm CE})/M_{\odot}}\, M_{\odot}\,{\rm yr}^{\rm
-1}.
  \end{equation}

If the CE does not exist, our treatment on the WD mass growth is
similar to that in \citet{HAN04} and \citet{MENG09}, i.e. our CEW
model reduces to the OTW model. (1) When $|\dot{M}_{\rm 2}|$ is
higher than $\frac{1}{2}\dot{M}_{\rm cr}$, hydrogen shell burning
is steady and no mass is lost from the system. The systems at this
phase may show the properties of SSSs. (2) When $|\dot{M}_{\rm
2}|$ is lower than $\frac{1}{2}\dot{M}_{\rm cr}$ but higher than
$\frac{1}{8}\dot{M}_{\rm cr}$, a very weak shell flash is
triggered but no mass is lost from the system. The systems at this
phase may show the properties of RNe. (3) When $|\dot{M}_{\rm 2}|$
is lower than $\frac{1}{8}\dot{M}_{\rm cr}$, the hydrogen-shell
flash is too strong to accumulate material on the surface of the
CO WD. Then, $\dot{M}_{\rm WD}$ is determined by
 \begin{equation}
 \dot{M}_{\rm WD}=\eta_{\rm He}\eta_{\rm
 H}|\dot{M}_{\rm 2}|,
  \end{equation}
where $\eta _{\rm H}$ and $\eta _{\rm He}$ are the mass
accumulation efficiencies for hydrogen burning and helium flashes,
respectively,
 \begin{equation}
\eta _{\rm H}=\left\{
 \begin{array}{ll}
 1, & \dot{M}_{\rm cr}\geq |\dot{M}_{\rm 2}|\geq\frac{1}{8}\dot{M}_{\rm
 cr},\\
 0, & |\dot{M}_{\rm 2}|< \frac{1}{8}\dot{M}_{\rm cr};
\end{array}\right.
\end{equation}
$\eta _{\rm He}$ is taken from \citet{KH2004}, and the treatment
of $\eta _{\rm He}$ is the same to that in \citet{MENG09}. We will
discuss the effects of $\eta _{\rm He}$ on the final results in
section \ref{subs:6.2}.

The model described above is still very simplistic and contains
several rather uncertain variables, such as the CE density that
has a major effect on the evolution of the system. We will discuss
the effects of these uncertainties in some detail in section
\ref{subs:6.1}.

We incorporated our model into the Eggleton stellar evolution code
and followed the evolution of the mass donor and the mass growth
of the accreting CO WD. We calculated more than 1100 WD+MS binary
sequences with different initial WD masses, initial secondary
masses and initial orbital periods, obtaining a rather dense grid
of models. The initial masses of donor stars, $M_{\rm 2}^{\rm i}$,
range from 1.8\,$M_{\odot}$ to 4.1\,$M_{\odot}$; the initial
masses of the CO WDs, $M_{\rm WD}^{\rm i}$, from 0.65\,$M_{\odot}$
to 1.20\,$M_{\odot}$; the initial orbital periods of binary
systems, $P^{\rm i}$, from the minimum value, where a zero-age
main-sequence (ZAMS) star just fills its Roche lobe, to $\sim
15$\,d, at which the companion star fills its Roche lobe at the
end of the HG. In this paper, we assume that a SN Ia occurs if the
mass of the CO WD reaches 1.378\,$M_{\odot}$, i.e.\ a value close
to the Chandrasekhar mass limit for non-rotating WDs
(\citealt{NTY84})\footnote{As rotating WDs may explode at a higher
mass than $M_{\rm WD}=1.378~M_{\odot}$, we continue our
calculations beyond this mass, assuming the same WD growth pattern
as for $M_{\rm WD}<1.378~M_{\odot}$ until $M_{\rm
WD}=1.45~M_{\odot}$ or until the mass-transfer rate falls beyond a
threshold value.}.

\section{BINARY EVOLUTION RESULTS}\label{sect:3}

\subsection{Binary evolution sequences for three examples in the
CEW model}\label{subs:3.1} For a WD + MS system, if the
mass-transfer rate, $|\dot{M}_{\rm 2}|$, exceeds the critical
value, $\dot{M}_{\rm cr}$, the WD will become a RG-like object and
fill and ultimately overfill its Roche lobe. Then, a CE forms. The
binary system will spiral in because of the frictional drag caused
by the CE. If the spiral-in process is very fast, the system could
merger and then no SN Ia occurs. Whether the binary system avoids
a merger fate in the CE phase is determined by the competition
between the mass-transfer timescale and the merger timescale, i.e.
the system may re-emerge from the CE phase if the merger timescale
is longer than the mass-transfer timescale; otherwise a merger is
unavoidable. In the following, we show some typical binary
evolution sequences in our CEW model. The evolution of the binary
system in the CEW model is similar to that of the OTW model
presented in \citet{HAN04}, but some of the details can be quite
different. In Figs.~\ref{ce0822} and \ref{ha0822} we present an
example to compare the differences between the CEW model and the
OTW model. Fig.~\ref{ce0822} shows the evolution of the key binary
parameters, including the merger timescale, as well as the
evolutionary tracks of the donor stars in the Hertzsprung-Russell
(HR) diagram, and the evolution of the orbital period. In this
example, the initial system has donor and WD masses of $M_{\rm
2}^{\rm i}=2.2\,M_{\odot}$ and $M_{\rm WD}^{\rm i}=0.8\,
M_{\odot}$, respectively, and an initial orbital period of
$\log(P^{\rm i}/{\rm d})=0.40$. The donor star fills its Roche
lobe in the Hertzsprung gap, i.e.\ the system experiences early
Case B RLOF.  The mass-transfer rate exceeds $\dot{M}_{\rm cr}$
soon after the onset of RLOF, leading to the formation of a CE,
where part of the CE material is lost from the surface of the CE.
The mass of the CE is always lower than $7\times10^{\rm
-4}\,M_{\odot}$. After about $1.1\times10^{\rm 6}$\,yr, the
mass-transfer rate has decreased below the critical accretion
rate, and the CE disappears. During the CEW phase, the CEW has a
similar effect as the OTW in the sense of balancing the
mass-transfer rate and the accretion rate of the WD, except that
the wind velocity from the CE surface is much lower than in the
OTW. When the mass-transfer rate drops below $\dot{M}_{\rm
  cr}$, but is still higher than $\frac{1}{2}\dot{M}_{\rm cr}$, mass
loss stops; as hydrogen shell burning is still stable, the WD
continues to gradually increase its mass. When the WD mass reaches
$M_{\rm WD}^{\rm SN}=1.378\,M_{\odot}$, the WD is assumed to
explode as a SN Ia. At this point, the mass of the donor is
$M_{\rm 2}^{\rm SN}=0.9682\,M_{\odot}$, and the orbital period is
$\log(P^{\rm SN}/{\rm d})=0.1076$. Note that in panel (3) of
Fig.~\ref{ce0822}, there is a jump in the mass-loss rate just
after $1.1\times 10^6\,$yr. This jump is caused by the different
helium accumulation efficiencies for different WD masses in the
model of \citet{KH2004}, which may also lead to a very small jump
in $\dot{M}_{\rm WD}$ although it is insignificant in panel (3) of
Figs.~\ref{ce0822} and ~\ref{ha0822} (see the treatment of
$\eta_{\rm He}$ in \citealt{MENG09} and the small jump of
$\dot{M}_{\rm WD}$ in Figure 1 of \citealt{MENGYANG10}). Panel (4)
of Fig.~\ref{ce0822} shows that the frictional luminosity between
the binary system and the CE is very low during the CE phase; as a
consequence, the merger timescale due to the friction in the CE is
much longer than the mass-transfer timescale, i.e.\ the CE does
not strongly affect the evolution of the binary parameters, and
the system avoids a merger in the CE phase.

The evolution of the binary system in the OTW model (Fig.
\ref{ha0822}) is quite similar to that of the CEW model
(Fig.~\ref{ce0822}), i.e.\  RLOF begins in the Hertzsprung gap and
the WD explodes during a phase where hydrogen shell burning is
stable. When the WD mass reaches $M_{\rm WD}^{\rm
SN}=1.378\,M_{\odot}$, the parameters of the binary system are
$(M_{\rm 2}^{\rm SN}, \log P^{\rm SN})=(0.8578, 0.1653)$. In
Fig.~\ref{ha0822}, the mass-loss rate from the system is the
difference between the mass-transfer rate and the mass-growth rate
of the WD. The steps in the mass-loss rate in panel (3) of Fig.
\ref{ha0822} are caused by the assumptions about the different
helium accumulation efficiencies for different WD masses.

However, many details are different in Figs.~\ref{ce0822} and
\ref{ha0822}. For example, compared to the final state of the
system in Figs.~\ref{ce0822} and \ref{ha0822}, a slightly more
massive final secondary and a shorter orbital period are found in
the CEW model. The differences are mainly caused by the different
treatment of the mass-growth rate of the CO WD when $|\dot{M}_{\rm
2}|$ is larger than $\dot{M}_{\rm cr}$. At this stage, the
mass-growth rate of the CO WD for the CEW model is always higher
than in the OTW model, i.e.\ the WD may increase its mass more
efficiently in the CEW model than in the OTW model when
$|\dot{M}_{\rm 2}|$ exceeds $\dot{M}_{\rm cr}$. Another reason
contributing to the differences is the friction between the CE and
the binary system which extracts orbital angular momentum from the
binary system, leading to a slightly shorter orbital period.  For
these reasons, the WD in the CEW model also explodes earlier than
in the OTW model by about $\sim 2\times10^{\rm 5}$\,yr.

\begin{figure}
\vspace{5mm}
    \includegraphics[angle=270,scale=.65]{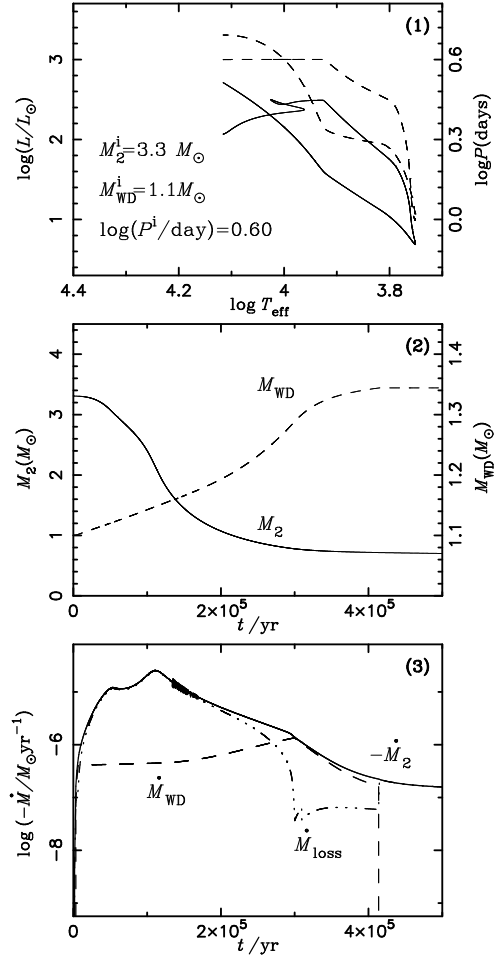}
\caption{Similar to Fig.~\ref{ha0822} but for different initial
binary parameters.} \label{ha0633}
\end{figure}

Actually, all the WD + MS systems that can explode as SNe Ia in
the OTW model will also explode in the CEW model, while some that
cannot explode in the OTW model do so in the CEW model.
Figs.~\ref{110633} and \ref{ha0633} show such an example. In this
example, both the WD and the donor are relatively massive. The
initial binary parameters in this case are $M_{\rm WD}^{\rm
i}=1.1\,M_{\odot}$, $M_{\rm 2}^{\rm i}=3.3\,M\odot$ and
$\log(P^{\rm i}/{\rm d})=0.60$. For both models, the donor star
fills its Roche lobe in the Hertzsprung gap. The mass-transfer
rate exceeds $\dot{M}_{\rm cr}$ soon after the onset of  RLOF
which results in the formation of the CE (or the onset of the
OTW). The WD then gradually grows its mass, but the final fate in
the two models is quite different, i.e.\ the WD mass in the CEW
model increases to $M_{\rm WD}^{\rm
  SN}=1.378\,M_{\odot}$ when $M_{\rm CE}=0.1559\,M_{\odot}$, while the WD does
not reach this mass in the OTW model before the mass-transfer rate
drops below $\frac{1}{8}\dot{M}_{\rm cr}$, below which novae are
assumed to prevent any further mass accumulation on the WD. The
different fate between the two models is again caused by the
different treatment of the mass-growth rate of the CO WD when
$|\dot{M}_{\rm 2}|>\dot{M}_{\rm cr}$. In addition, because of the
effect of the CE on the orbital period, the moment at which the
mass-transfer rate drops below $\dot{M}_{\rm cr}$ is slightly
delayed, which may also be responsible for the different fate
between the CEW model and the OTW model. Moreover, due to the
existence of the CE, the mass-growth rate of the WD may still
maintain a high value even when $|\dot{M}_{\rm 2}|<\dot{M}_{\rm
WD}$.

\begin{figure*}
\vspace{5mm}
    \includegraphics[angle=270,scale=.65]{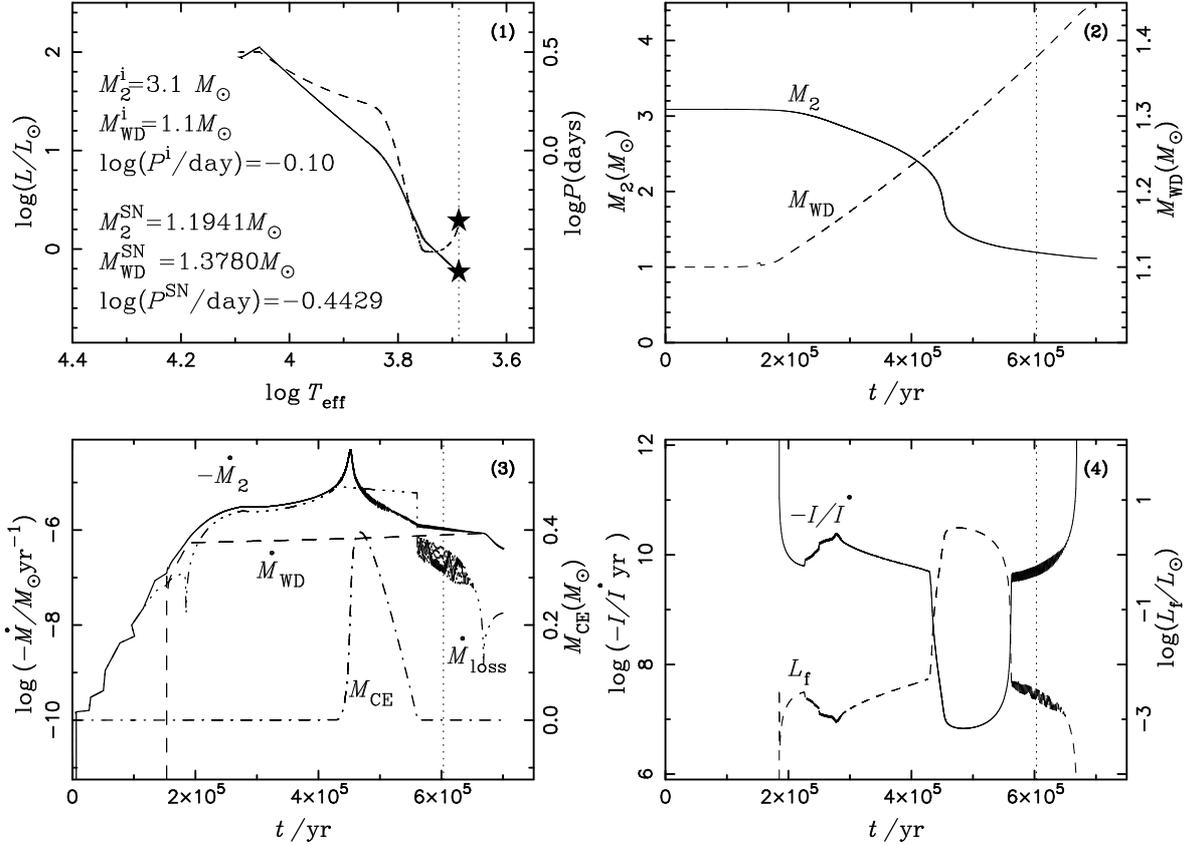}
\caption{Similar to Fig.~\ref{ce0822} for different initial binary
parameters, illustrating a relatively extreme case.} \label{delay}
\end{figure*}

As Figs.~\ref{ce0822} and \ref{110633} show, the CE mass is much
larger in the simulations in Fig.~\ref{110633} than in Fig.
\ref{ce0822}. At the moment when $M_{\rm WD}^{\rm
  SN}=1.378\,M_{\odot}$, the CE is still as massive as
0.1559\,$M_{\odot}$ (see panel 3 of Fig.~\ref{110633}). So, the
mass-growth rate of the CO WD can maintain a high value even when
$|\dot{M}_{\rm 2}|$ is lower than $\dot{M}_{\rm WD}$. Furthermore,
the frictional luminosity in Fig.~\ref{110633} is also much larger
than that in Fig.~\ref{ce0822} because of the more massive CE.
Even though the frictional merger timescale in Fig.~\ref{110633}
is shorter than that in Fig.~\ref{ce0822}, it is still longer than
the mass-transfer timescale, i.e.\ the effect of the CE on the
binary evolution may still be small even for a CE of
$\sim0.7\,M_{\odot}$. Actually, if the frictional luminosity is
lower than several $10^{\rm 2}\,L_{\odot}$ to $10^{\rm
3}\,L_{\odot}$, the binary system may survive from the CE phase
irrespective of the CE mass.

Fig.~\ref{delay} illustrates a more extreme case where both the
donor and the WD are relatively massive, but the initial orbital
period is shorter than that in Fig.~\ref{110633}. The initial
binary parameters in this case are $M_{\rm WD}^{\rm i}=1.1\,
M_{\odot}$, $M_{\rm 2}^{\rm i}=3.1\,M_{\odot}$ and $\log(P^{\rm
i}/{\rm d})=-0.10$. For this short initial orbital period, the
donor star starts to fill its Roche lobe on the MS. In the first
$\sim4\times10^{\rm 5}$\,yr, the donor loses about
$0.5\,M_{\odot}$ after the onset of RLOF. At this stage, mass
transfer almost becomes dynamically unstable, and hence the
mass-transfer rate increases sharply, as does the CE mass. Within
the following $\sim5\times10^{\rm 4}$\,yr, the companion loses
about 1\,$M_{\odot}$. The mass-transfer rate drops only after the
mass ratio has been reversed. The binary parameters at the
explosion are $(M_{\rm 2}^{\rm SN}, \log P^{\rm SN})=(1.1941,
-0.4429)$. For an even larger initial donor mass, e.g.\ $M_{\rm
2}^{\rm i}=3.2\,M_{\odot}$, our calculations show that mass
transfer becomes unstable, and such systems may experience a
delayed dynamical instability (\citealt{HW87}), i.e.\ the system
may merge completely
(but also see \citealt{HAN06}). Notice that the oscillations in
some curves in Fig.~\ref{delay} are numerical artifacts caused by
the high mass-loss rate of the donor for this binary system.

\begin{figure*}
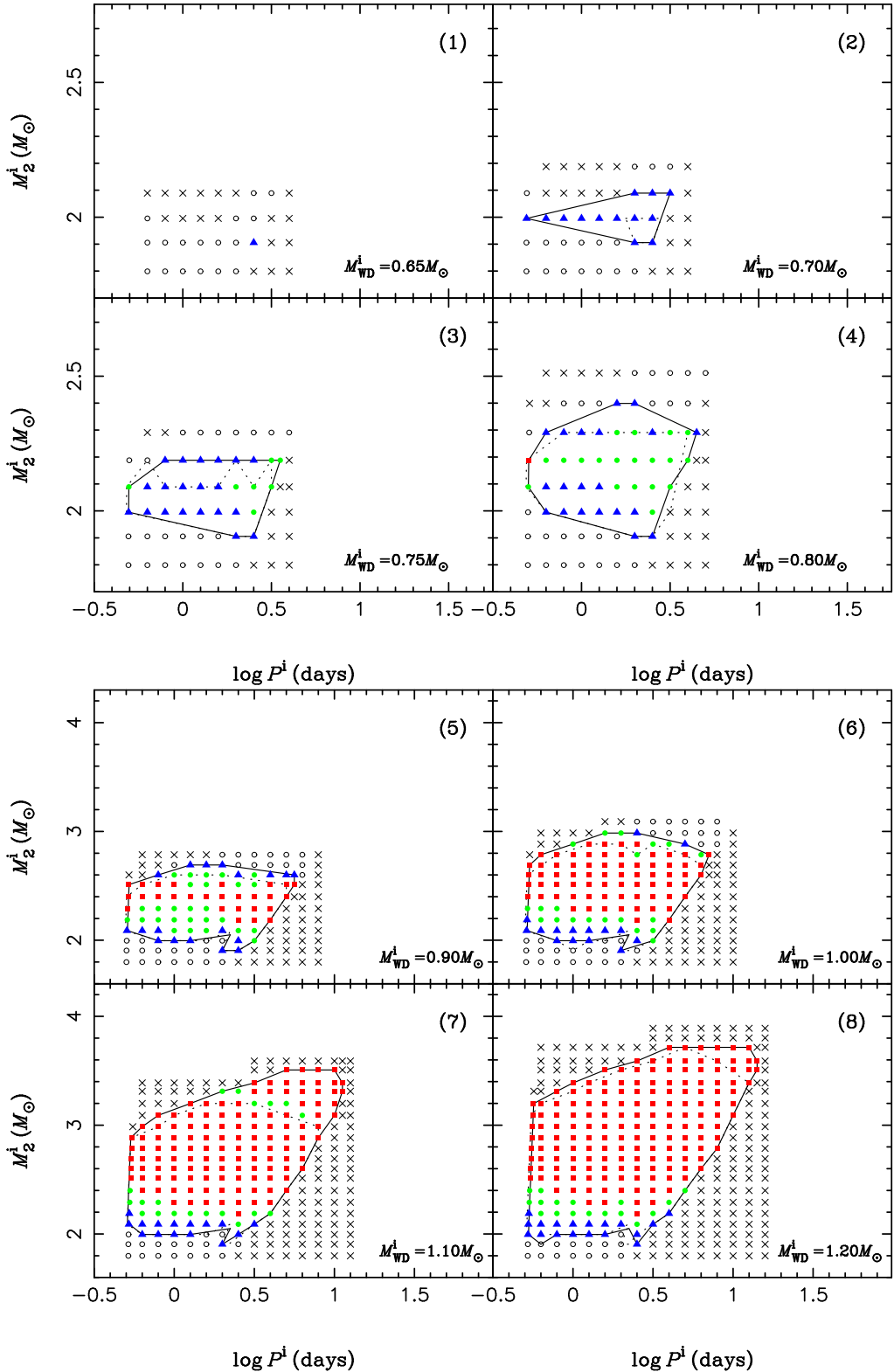

    \includegraphics[angle=270,scale=.60]{grid2.ps}
    \includegraphics[angle=270,scale=.60]{grid1.ps}

\caption{Final outcomes of the binary evolution calculations in
the CEW model in the initial orbital period -- secondary mass
($\log P^{\rm i}, M_{\rm 2}^{\rm i}$) plane, where $P^{\rm i}$ is
the initial orbital period, and $M_{\rm 2}^{\rm i}$ is the initial
mass of the donor star (for different initial WD masses as
indicated in each panel). Filled squares indicate SN Ia explosions
during a CE phase ($M_{\rm CE}>0$). Filled circles denote that SN
Ia explosions occur in the SSS phase ($\dot{M}_{\rm cr}\geq
|\dot{M}_{\rm 2}|\geq \frac{1}{2}\dot{M}_{\rm cr}$ and $M_{\rm
CE}=0$), while filled triangles denote that SN Ia explosions occur
in the RN phase ($\frac{1}{2}\dot{M}_{\rm cr}> |\dot{M}_{\rm
2}|\geq \frac{1}{8}\dot{M}_{\rm cr}$ and $M_{\rm CE}=0$). Open
circles indicate systems that experience nova explosions,
preventing the CO WD from reaching 1.378\,$M_{\odot}$
($|\dot{M}_{\rm 2}|<\frac{1}{8}\dot{M}_{\rm cr}$ and $M_{\rm
CE}=0$), while crosses show the systems that are unstable to
dynamical mass transfer. The solid curves show the contours of the
parameter space leading to SNe Ia for the CEW model, while, for
comparison, the dotted curves show the contours from the OTW model
(taken from \citealt{MENG09}).} \label{grid}
\end{figure*}

\subsection{Final outcomes of the binary evolution calculations}\label{subs:3.2}
To clearly show the difference between our results and previous
ones in literatures, we show the final outcomes of all the binary
evolution calculations in an initial orbital period -- secondary
mass ($\log P^{\rm i}, M_{\rm 2}^{\rm i}$) plane
(Fig.~\ref{grid}).  As the figure shows, CO WDs may reach a mass
of 1.378\,$M_{\odot}$ while the system is still in the CE phase
(filled squares) or after the CE evolution has ended, where they
can be either in stable (filled circles) or weakly unstable
(filled triangles) hydrogen-burning phases.  Systems re-emerging
from the CE phase may show the properties of SSSs if hydrogen
burning on the WD is stable or those of RNe for weakly unstable
hydrogen burning.  All these systems are probably progenitors of
SNe Ia. Because of dynamically unstable mass transfer or strong
hydrogen shell flashes, many CO WDs do not increase their masses
to 1.378\,$M_{\odot}$. As shown in Figs.~\ref{110633} and
\ref{ha0633}, in comparison to the OTW model, the longer timescale
for the CEW phase and the higher mass-growth rate during this
phase result in a larger increase of the CO WD mass in this phase.
Consequently the timescale from the end of the CEW to the
explosion (where the systems may appear as SSSs or RNe) is
shorter, or the CO WD, which would not increase to
1.378\,$M_{\odot}$ in the OTW phase, may reach 1.378\,$M_{\odot}$
during the CEW phase (the system then will not appear as a SSS or
RN system). This means that a given system is less likely to be in
the SSS/RN phase at the SN explosion in the CEW model
(\citealt{MENG09,MENG09c}). Based on the supersoft X-ray flux in
elliptical galaxies, \citet{GB10} concluded that no more than 5\%
of SNe Ia in early-type galaxies can be produced by mass-accreting
white dwarfs from the SD scenario. According to the OTW model,
\citet{MENGYANG11b} found that the mean relative duration of the
SSS phase for an accreting WD is about 5\%. However, based on the
small number of SSSs observed in some galaxies such as M31,
\citet{MAOZ14} argued that no more than 1\,\% of the WD's growth
time is spent in the SSS phase. This is an order of magnitude less
than was found by \citet{MENGYANG11b}. The results here should
help to resolve the conflict between \citet{MAOZ14} and
\citet{MENGYANG11b} since the CE phase in the CEW model lasts
longer than the OTW phase. Moreover, even if the binary system
were in a SSS phase, the supersoft X-ray flux would also be
strongly reduced by the CSM that forms because of the CEW or the
OTW, but the CSM from our CEW model would be more efficient in
suppressing the supersoft X-ray flux because of the relatively
high CSM density due to the much lower wind velocity
(\citealt{NIELSEN13}; \citealt{WHELLER13}).


In Fig.~\ref{grid} we present contours of the initial parameters
leading to SNe Ia for different WD masses, indicating the final
state of the system, while Fig.~\ref{cour} combines the contours
in one summary plot. The left boundaries of the contours are
determined by the radii of ZAMS stars, i.e.\ correspond to systems
that start RLOF on the ZAMS; systems beyond the right boundaries
experience dynamically unstable mass transfer at the base of the
red-giant branch (RGB). The upper boundaries are determined by
systems that experience a delayed dynamical instability. For the
systems above the lower boundaries, the mass-transfer rate is
larger than $\frac{1}{8}\dot{M}_{\rm cr}$, which prevents the
occurrence of strong hydrogen shell flashes, while, at the same
time, the secondaries can provide enough material to feed the
growth of the CO WDs to allow it to reach 1.378\,$M_{\odot}$. The
figure shows that the initial WD mass significantly affects the
upper boundary but not the lower boundary. The upper boundary
moves to lower masses as the initial WD mass decreases, causing a
shrinking of the initial parameter space that leads to SNe Ia with
a decrease of the WD mass; at $M_{\rm WD}^{\rm i}=0.65\,M_{\odot}$
the contour collapses to a point.

\subsection{Comparison with the OTW model}\label{subs:3.3}
Fig.~\ref{grid} also shows the contours of initial parameters
leading to SNe Ia from the OTW model for comparing the two model.
The shapes of the contours are quite similar to each other, except
that the upper boundary of the CEW model is higher than that in
the OTW model and that the maximum initial orbital period for the
CEW model can be longer; this means that mass transfer between the
binary components is somewhat more stable for the CEW model than
for the OTW model. This phenomenon is mainly caused by the
different treatment of the mass-growth rate of the CO WD; i.e.\
the mass-growth rate during the CE evolution phase in the CEW
model is higher than that during the OTW phase in the OTW model.
So, for a given binary system, the WD mass (the mass ratio,
$M_{\rm 2}/M_{\rm WD}$) at the same evolutionary stage for the CEW
model is slightly higher (smaller) than in the OTW model, which
leads to relatively more stable mass transfer for the CEW model.
Especially in the right-upper part of the allowed parameter space
for the CEW model, the systems will experience nova explosions,
preventing the CO WD from reaching 1.378\,$M_{\odot}$ for the OTW
model, while the WD will reach 1.378\,$M_{\odot}$ in the CE phase
for the CEW model. This difference arises from the fact that the
CE acts as a mass reservoir which can continuously feed material
to the WD as long as the CE exists, even when $|\dot{M}_{\rm
2}|<\dot{M}_{\rm cr}$ as shown in Fig.~\ref{110633}.

Another difference is the minimum initial mass of CO WDs, $M_{\rm
  WD}^{\rm min}$, that can lead to SNe Ia. Previous studies showed
that the minimum initial mass may be as low as $0.67\,M_{\odot}$
for $Z=0.02$ based on the OTW model (\citealt{LAN00};
\citealt{HAN04}), and \citet{MENG09} found that $M_{\rm WD}^{\rm
min}$ strongly depends on metallicity. Here, the minimum mass of
CO WDs for $Z=0.02$ is $0.65\,M_{\odot}$, slightly lower than that
for the OTW model, which is also a direct consequence of the
different treatment of the mass-growth rate of the WDs.

These differences between the CEW model and the OTW model already
indicate that the birth rate of SNe Ia in the CEW model should be
somewhat higher than in the OTW model (see section \ref{sect:5}).

\subsection{The final state of the systems}\label{subs:3.4}
In this section, we examine the final state of the binary systems
and the properties of the companions at the time of the supernova
explosion (assumed to occur when the WD mass has reached $M_{\rm
  WD}=1.378\,M_{\odot}$); this may be helpful for identifying SN Ia
progenitor systems or for finding surviving companions in
supernova remnants.

\begin{figure}
    \includegraphics[angle=270,scale=.35]{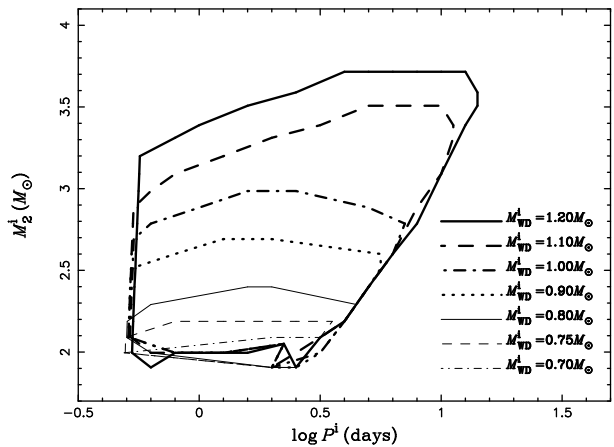}
\caption{The contours of initial parameters in the ($\log P^{\rm
i},M_{\rm 2}^{\rm i}$) plane for different initial WD masses for
which SNe Ia are expected. For $M_{\rm WD}^{\rm
i}=0.65\,M_{\odot}$, the contour shrinks to a point and is not
included in the figure. } \label{cour}
\end{figure}

\begin{figure*}
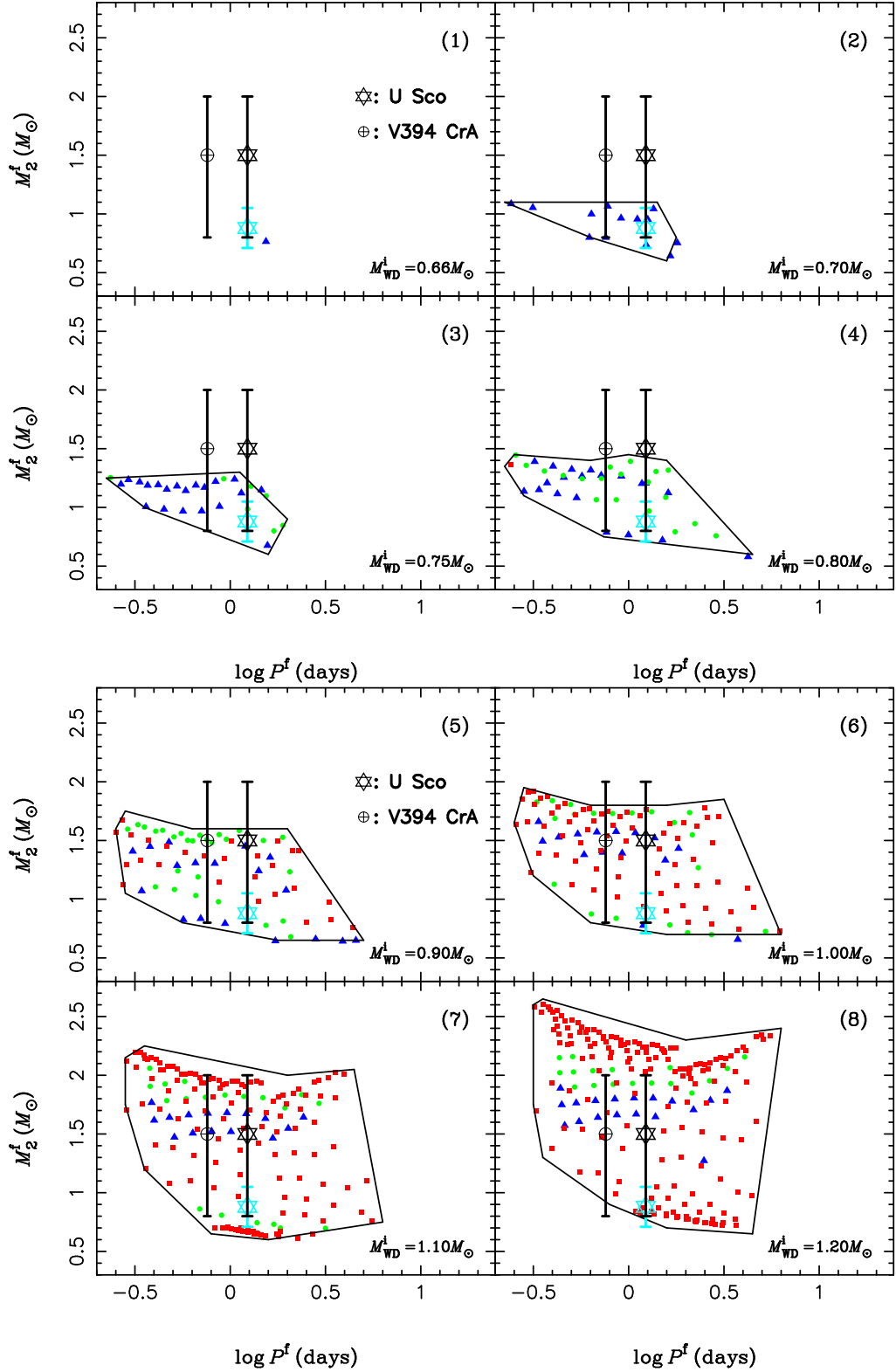

    \includegraphics[angle=270,scale=.60]{final2.ps}
    \includegraphics[angle=270,scale=.60]{final1.ps}
\caption{Similar to Fig.~\ref{grid} but for the final state of the
binary systems in the CEW model in the final orbital period --
secondary mass ($\log P^{\rm f}, M_{\rm 2}^{\rm f}$) plane, where
$P^{\rm f}$ is the final orbital period, and $M_{\rm 2}^{\rm f}$
is the final mass of the donor star (for different initial WD
masses as indicated in each panel) at the moment of $M_{\rm
WD}=1.378~M_{\odot}$. Filled squares indicate SN Ia explosions
during a CE phase ($M_{\rm CE}>0$). Filled circles denote that SN
Ia explosions occur in the SSS phase ($\dot{M}_{\rm cr}\geq
|\dot{M}_{\rm 2}|\geq \frac{1}{2}\dot{M}_{\rm cr}$ and $M_{\rm
CE}=0$), while filled triangles indicate that SN Ia explosions
occur in the RN phase ($\frac{1}{2}\dot{M}_{\rm cr}> |\dot{M}_{\rm
2}|\geq \frac{1}{8}\dot{M}_{\rm cr}$ and $M_{\rm CE}=0$). The
solid curves show the contours of the parameter space. Two
recurrent novae are indicated by a hexagram for U Sco and an Earth
symbol for V394 CrA, where the position of the black hexagram is
based on the the model simulation by \citet{HAC00a,HAC00b}, while
the bright blue hexagram taken from the dynamical estimate
(\citealt{THOROUGH01}).} \label{fin}
\end{figure*}

\subsubsection{Final contour}\label{subs:3.4.1}
\begin{figure}
    \includegraphics[angle=270,scale=.35]{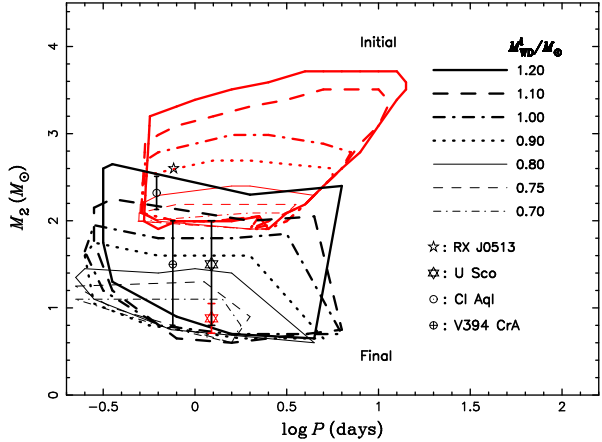}
\caption{Parameter regions producing SNe Ia in the ($\log P-M_{\rm
    2}$) plane for WD + MS systems with different initial WD masses in
  the CEW model. In the WD + MS systems inside the region enclosed by
  the red curves (labelled ``initial''), the white dwarf is able to
  grow to a mass of 1.378\,$M_{\odot}$, where a SN Ia explosion is
  assumed to occur.  The final state of the WD + MS systems is
  enclosed by the black curves (labelled ``final''). A supersoft X-ray
  source, RX J0513.9-6951 (open star) is also plotted; its orbital
  period is 0.763\,d (\citealt{PAKULL93}). Three recurrent novae are
  indicated by a hexagram for U Sco (with a period of 1.2306\,d;
  \citealt{SR95}; \citealt{HAC00a, HAC00b}), a solar symbol for CI Aql
  (with an orbital period of 0.6184\,d; \citealt{MH95};
  \citealt{SAHMAN13}) and an Earth symbol for V394 CrA (with an
  orbital period of 0.7577\,d; \citealt{SCHAEFER90}), where
    the black hexagram assumes the companion mass from the model
    simulation by \citet{HAC00a,HAC00b}, while the red hexagram takes
    it from the dynamical estimate
    (\citealt{THOROUGH01}).} \label{final}
\end{figure}

Fig.~\ref{fin} shows the final states of binary systems leading to
SNe Ia in the ($\log P^{\rm f}-M_{\rm 2}^{\rm f}$) plane when
$M_{\rm WD}=1.378~M_{\odot}$ for different initial WD masses,
while Fig.~\ref{final} compares the initial and final contours of
these  systems. Generally, for a system with given initial
parameters, the companion is still massive enough to maintain a
high  mass-transfer rate at the moment when $M_{\rm
WD}=1.378~M_{\odot }$. For example, many systems are still in the
CEW phase at this moment; this means that the WDs in these systems
would continue to increase their masses if the WDs did not explode
at this stage. Even for the systems exploding in the SSS or RN
phase, the companions may still be massive enough to maintain a
high mass-transfer rate, so that the WDs could continue to grow in
mass if they did not explode when $M_{\rm WD}=1.378~M_{\odot }$
(see Figs.~\ref{ce0822} and \ref{fin}).  In other words, if the WD
of a system is  slightly less massive than 1.378 $M_{\odot}$ and
its companion mass and orbital period are also located in the
final contours for SNe Ia, it is still very  likely for the WD to
continue to increase its mass to 1.378 $M_{\odot}$.
Fig.~\ref{final} shows that the positions of the  final contours
are below those of the initial contours; the secondary always has
a mass less than 2.6\,$M_{\odot}$, but can be as low as
$0.6\,M_{\odot}$. Almost all models have a final orbital period
between 0.22\,d and 6.5\,d; these periods are generally shorter
than in the OTW model (see \citealt{MENG09c}). Similar to the
initial contours, the parameter range of the final contours
shrinks with decreasing initial WD mass and disappears when
$M_{\rm WD}=0.65\,M_{\odot}$. A SSS (RX J0513.9-6951) and three
RNe (CI Aql, U  Sco and V394 CrA) are also indicated in
Fig.~\ref{final} and two RNe (U  Sco and V394 CrA) in
Fig.~\ref{fin}.  These objects have been considered good
candidates for progenitor systems of SNe Ia as they contain
massive WDs and have short orbital periods (\citealt{PAR07}).
Numerical simulations provide further support for the SN Ia
progenitor connection (see below). RX  J0513.9-6951 (open star)
with $M_{\rm WD}=1.30\,M_{\odot}$  (\citealt{HK03b}) is located
within the initial contour for SN Ia progenitors. If the WD is  a
CO WD, it would make an excellent candidate for a SN Ia
progenitor, as already suggested in \citet{HK01} and
\citet{HK03b}. For CI Aql, \citet{SAHMAN13} recently analyzed
time-resolved, intermediate-resolution spectra during quiescence
and found a WD mass of $1.00\pm0.14\,M_{\odot}$ and a secondary
mass of $2.32\pm0.19\,M_{\odot}$. The position of CI Aql is also
perfectly placed within the initial contours for SNe Ia. Even
considering the uncertainty in the WD mass, our model suggests
that CI Aql is another  excellent SN Ia progenitor candidate
system.

The dynamical mass estimates for the binary components in U Sco
are $M_{\rm WD}=1.55\pm0.24\, M_{\odot}$ and $M_{\rm
2}=0.88\pm0.17\,M_{\odot}$ (\citealt{THOROUGH01}), while model
simulations show U Sco to be a WD + MS system with $M_{\rm
WD}=1.37\,M_{\odot}$ and $M_{\rm 2}=1.5\, M_{\odot}$, although a
mass between $0.8\,M_{\odot}$ and $2.0\,M_{\odot}$ is also
acceptable (\citealt{HAC00a,HAC00b}, see also the binary evolution
model in \citealt{POD03}). Both the dynamical estimates and model
simulations indicate that the mass of the WD in U Sco is quite
close to 1.378 $M_{\odot}$. Figs.~\ref{fin} and \ref{final} that
show that U Sco is well located within the final contours for SNe
Ia, irrespective of whether the companion mass is taken from model
simulations or dynamical estimates; this suggests that the
companion has enough material to maintain a high mass-transfer
rate to increase the WD mass to 1.378\,$M_{\odot}$. Therefore, our
model supports the previous suggestion that, if the WD in U Sco is
a CO WD, it is an excellent candidate for a SN Ia progenitor.
However, whether or not the massive WD is a CO or a ONeMg one is
still under debate (\citealt{MASON11}; \citealt{KH2012}). Further
effort, both on the observational and the theoretical side, is
clearly needed to decide the fate of U Sco. The WD mass of V394
CrA has been estimated to be 1.37\,$M_{\odot}$, while the mass of
the companion is still unclear. The best-fit companion mass for
this recurrent nova system is 1.50\,$M_{\odot}$, while a mass
between $0.8\,M_{\odot}$ and $2.0\,M_{\odot}$ is still acceptable
(\citealt{HK00,HK03a}; \citealt{HKS03}). In either case, it is
still very likely that the WD can increase its mass to
1.378\,$M_{\odot}$. Assuming a CO composition for the WD of V394
CrA, our results confirm that it provides an excellent candidate
for a SN Ia progenitor. In any case, further observations are
necessary to determine the masses of the WD and the secondary in
V394 CrA.

In addition, as the well candidates of the progenitors of SNe Ia,
U Sco and V394 CrA provide an opportunity to examine the
progenitor models of SNe Ia in a detailed way, i.e. whether or not
the two RNe may locate in the RN region for SNe Ia predicted by
any successful single-degenerate model. In Fig.~\ref{fin},
although the expected regions for the CEW, SSS and phases
overlapped each other, the RN region predicted by our CEW model
may explain the positions of U Sco and V394 CrA in the ($\log
P^{\rm f}-M_{\rm 2}^{\rm f}$) plane. According to Fig.~\ref{fin},
both U Sco and V394 CrA are likely to come from systems with an
initial WD mass between $1.0~M_{\odot}$ and $1.1~M_{\odot}$ if the
best-fit companion masses from model simulations are taken, while
the dynamical estimate for U Sco favors a system with $M_{\rm
WD}^{\rm i}\leq0.9~M_{\odot }$.

\subsubsection{Mass and orbital velocity}\label{subs:3.4.2}
\begin{figure}
    \includegraphics[angle=270,scale=.35]{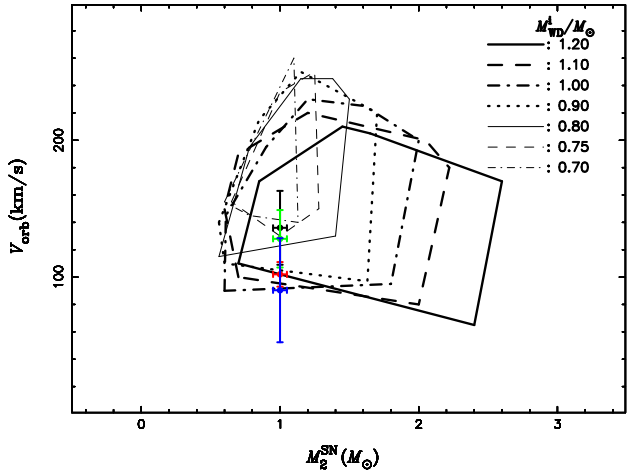}
\caption{Parameter space for the companion mass and its orbital
  velocity when $M_{\rm WD}=1.378\,M_{\odot}$ for different initial WD
  masses in the CEW model. The cross represents the potential
  companion of Tycho's supernova progenitor, Tycho G, and the length
  of the cross indicates the typical observational errors, where the
  space velocities for Tycho G are taken from \citet{RUI04} (black), \citet{GONZA09} (green),
 \citet{KERZENDORF09} (blue) and \citet{BEDIN14} (red).} \label{m2vorb}
\end{figure}

In Fig.~\ref{m2vorb} we show the parameter regions for the
companion mass and its orbital velocity when $M_{\rm
WD}=1.378\,M_{\odot}$ for different initial WD masses. The ranges
of companion mass and orbital velocity here are very similar to
those shown in \citet{HAN08} and \citet{MENGYANG10b} although
their results were based on the OTW model. At the time of the
supernova, the companion mass lies between 0.6\,$M_{\odot}$ and
2.6\,$M_{\odot}$ and the orbital velocity between 60\,${\rm
km\,s^{\rm -1}}$ and 260\,${\rm km\,s^{\rm
    -1}}$. The impact of the supernova ejecta with the envelope of the
companion may strip off some hydrogen-rich material from the
surface of the companion. At the same time, the companion may
receive a velocity kick, i.e.\ the final space velocity of a
surviving companion in a supernova remnant is $v_{\rm
s}=\sqrt{v_{\rm orb}^{\rm 2}+v_{\rm
    kick}^{\rm 2}}$ (\citealt{MAR00}; \citealt{MENG07};
\citealt{PAKMOR08}; \citealt{LIUZW12}). Generally, the
stripped-off material is much less massive than the companion, and
the kick velocity is much smaller than the orbital velocity. In
any case, the companion mass and its orbital velocity here should
be taken as upper and lower limits for the mass and the space
velocity of the surviving companion in a supernova remnant,
respectively. In the figure, we also plotted the parameters for
the potential surviving companion of Tycho's supernova (Tycho G,
\citealt{RUI04}; \citealt{GONZA09}; \citealt{BEDIN14}). Although
it is presently unclear whether Tycho G actually is the surviving
companion (see the discussion in section \ref{subs:3.4.3}), its
claimed parameters would be compatible with the range of the
companion mass and its orbital velocity presented here.

\begin{figure}
    \includegraphics[angle=270,scale=.35]{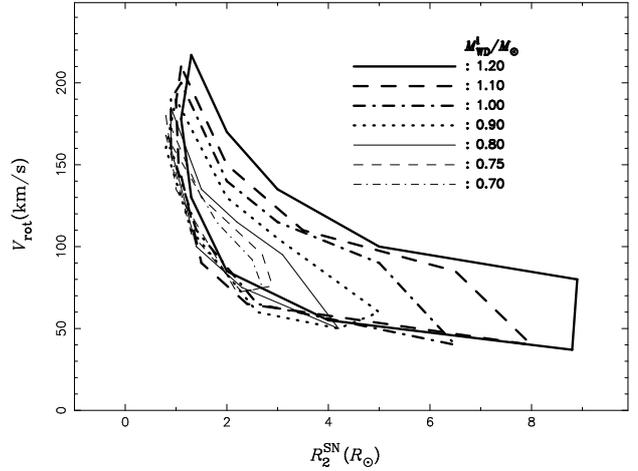}
\caption{Parameter regions for the companion radius and its
equatorial
  rotational velocity when $M_{\rm WD}=1.378\,M_{\odot}$
  for different initial WD masses in the CEW model.}
\label{rvrot}
\end{figure}

\subsubsection{Companion radius and rotational velocity}\label{subs:3.4.3}
In Fig.~\ref{rvrot}, we present the parameter regions for the
companion radius and its equatorial rotational velocity when
$M_{\rm WD}=1.378\,M_{\odot}$ for different initial WD masses. In
our calculations, we did not keeptrack of the companion radius as
it can be obtained by some simple assumptions. We assume that the
companion radius is equal to its Roche-lobe radius, i.e. the
companion is filling its Roche lobe at the moment of the supernova
explosion, and the donor's radius will not be far away from that
shown in Fig.~\ref{rvrot}. The equatorial rotation velocity is
calculated assuming that the companion star co-rotates with the
orbit. The figure shows that the rotational velocity of the
surviving companions ranges from 40\,km\,s$^{\rm -1}$ to
220\,km\,s$^{\rm -1}$; this means that, in many cases, the
companions' spectral lines should be noticeably broadened if the
rotation of the star is not affected by the impact of the
supernova ejecta. We note that the upper limit of the rotational
velocity here is higher than that based on the OTW model in
\citet{HAN08} (170\,km\,s$^{\rm -1}$) since the final orbital
periods tend to be shorter in the CEW model than in the OTW model
(see Figs.~\ref{ce0822} and \ref{ha0822}). The companion radius
lies between 0.5\,$R_{\odot}$ and 9\,$R_{\odot}$. In general, the
companion stars with larger radii have lower rotational
velocities. This figure may be helpful for identifying the
companion nature of a potential surviving candidate in a supernova
remnant. However, although the radius of Tycho G
($1-3\,R_{\odot}$) matches well with our calculations, its
rotational velocity is much smaller than predicted here; this is
the reason why \citet{KERZENDORF09} have strongly argued against
Tycho G being the surviving companion of Tycho's supernova. Here,
we do not consider the impact of the supernova ejecta on the
companion. If the companion expands significantly because of the
impact, its rotational velocity will be significantly reduced.
Indeed, as argued by \citet{KERZENDORF09} (also see
\citealt{MENGYANG11}), it is possible in principle to explain the
observed rotational velocity in Tycho G if the companion's
envelope was almost completely stripped by the supernova impact
and the surface layers we see now had to expand from regions near
the core. These expectations, that follow from simple
angular-momentum conservation considerations, have been confirmed
by  numerical simulations (e.g.\ \citealt{PANK12};
\citealt{LIUZW13}), though we note that none of the simulations in
\citet{PANK12} can explain the observed properties of Tycho G as
none can fit the observed gravity ($\log g$) and the observed
rotational velocity simultaneously.

\begin{figure}
    \includegraphics[angle=270,scale=.35]{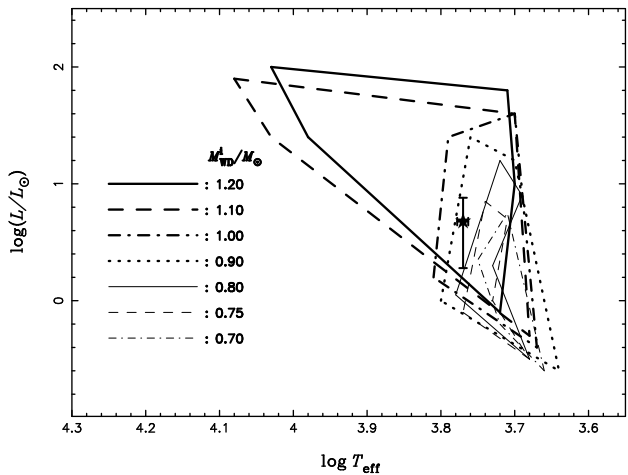}
\caption{The parameter regions for the luminosity and the
temperature of
  the companion when $M_{\rm WD}=1.378\,M_{\odot}$ for
  different initial WD masses. The star represents the potential surviving
  companion star of Tycho's supernova, Tycho G, where the bar represents
  the observational error (\citealt{RUI04}).}
\label{lumtef}
\end{figure}

\subsubsection{Luminosity and effective temperature}\label{subs:3.4.4}
Fig.~\ref{lumtef} shows the parameter regions for the luminosity
and the temperature of the companion when $M_{\rm
WD}=1.378\,M_{\odot}$ for different initial WD masses.  The
luminosity ranges from $0.25\,L_{\odot}$ to $100\,L_{\odot}$, and
the effective temperature from $4470$\,K to $12600$\,K. The range
of effective temperatures is slightly larger than in the OTW model
(\citealt{HAN08}), which again is a direct consequence of the
different treatment of the mass-growth rate of the WD, leading to
a larger initial parameter space in the initial secondary mass --
orbital period plane.

The luminosity and the effective temperature of a companion will
generally be affected by the impact of the supernova ejecta
(\citealt{MAR00}; \citealt{POD03}; \citealt{SHAPPEE13}); hence
Fig.~\ref{lumtef} can only give an indication for the initial
conditions when exploring the evolution and appearance of a
companion star during the post-impact re-equilibration phase when
the star is out of thermal equilibrium. \citet{POD03} calculated
the luminosity evolution of a subgiant star of 1\,$M_{\odot}$
after being hit by the ejecta of a SN Ia and found that the
luminosity evolution was initially much faster than the
Kelvin-Helmholtz timescale of the pre-SN subgiant since the
re-equilibration timescale is determined by the thermal timescale
of the outer layer of the star, which can be many orders of the
magnitude shorter.  Depending on the amount of energy deposited,
the luminosity of the companion may be either significantly
overluminous or underluminous ($0.1\,L_{\odot}$ --
$200\,L_{\odot}$) 440\,yr after the SN explosion. \citet{PANK12b}
found that the evolution of the remnant star strongly depends not
only on the amount of energy deposited from the explosion but also
on the depth of the energy deposition and that the luminosity of
the remnant could be close to that of Tycho G $\sim 500$\,yr after
the explosion (consistent with the conclusions of
\citealt{POD03}). \citet{SHAPPEE13} calculated the future
evolution of the companion by injecting $(2-6)\times10^{\rm
47}$\,erg of energy into the stellar-evolution model of a
1\,$M_{\odot}$ donor star and found that the companion becomes
significantly more luminous ($10-10^{\rm 3}\,L_{\odot}$) for a
long period of time ($10^{\rm
  3}-10^{\rm 4}$\,yr) due to the Kelvin-Helmholtz collapse of the
envelope. These simulations and our binary evolution calculations
partly share the luminosity range, especially for the results in
\citet{POD03}. The potential surviving companion of Tycho's
supernova is well located within the contour of Fig.~\ref{lumtef}.
However, as we discussed in section \ref{subs:3.4.3}, no model to
date can reproduce $\log g$, the rotation velocity and the
luminosity of Tycho G simultaneously. Observations are also not
fully conclusive. While \citet{BEDIN14} in their re-analysis
conclude that the proper motion and chemical abundance of Tycho G
are consistent with it being a surviving companion,
\citet{XUEZC15}, measuring the exact explosion site of Tycho's
supernova, found that Tycho G is outside the 1$\sigma$ error
region (7.3 arcsec) at very high significance. Therefore Tycho G
may just be a Milky Way thick-disc star that happens to pass in
the vicinity of the supernova remnant (\citealt{FUHRMANN05}; but
also see \citealt{BEDIN14} for a different view).

\begin{figure}
    \includegraphics[angle=270,scale=.35]{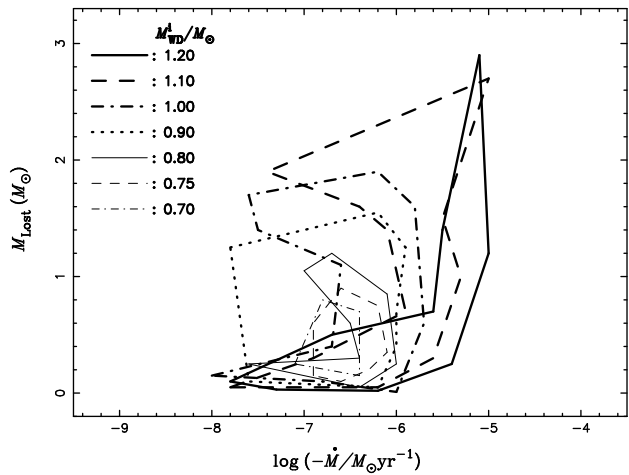}
\caption{The parameter regions for the mass-lose rate and the
total amount of material lost from the system when $M_{\rm
WD}=1.378\,M_{\odot}$ for different initial WD masses, where the
CE mass is included in the amount of material lost from the
system.} \label{mlost}
\end{figure}

\begin{figure}
    \includegraphics[angle=270,scale=.35]{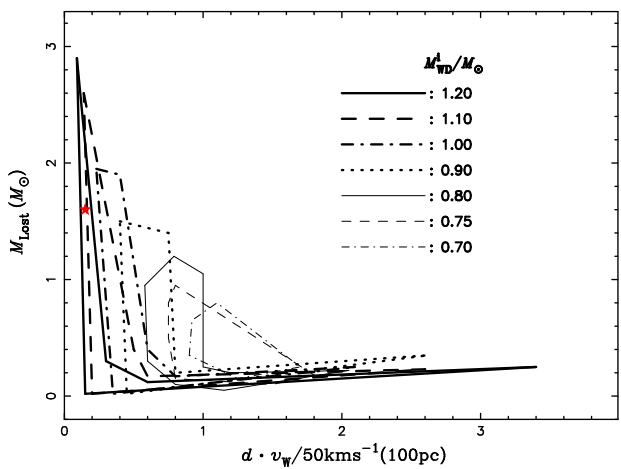}
\caption{The parameter regions for the total amount of material
lost from the system and the maximum distance that the lost
material may reach by the time when $M_{\rm WD}=1.378\,M_{\odot}$
for different initial WD masses; here the CE mass is included in
the amount of material lost from the system. The red star
represents the Galactic SN remnant RCW 86 (\citealt{BROERSEN14}).}
\label{mlpc}
\end{figure}

\subsubsection{Mass loss}\label{subs:3.4.5}
Fig.~\ref{mlost} shows the mass-lose rate from the system and the
total amount of material lost when $M_{\rm WD}=1.378\,M_{\odot}$.
Here, the CE mass is included in the material lost since the CE
material may generally be lost eventually (see the further
discussion in section \ref{subs:6.1}) or provide a mass reservoir
to form a circumbinary disk. This assumption for the material lost
is reasonable for about 99\% of all SNe Ia, because the CE mass is
so small (less massive than $\sim 10^{\rm -3}~M_{\odot }$) for
most SNe Ia exploding in the CE phase.
Only for the cases of $M_{\rm WD}^{\rm i}\geq1.1~M_{\odot}$, may
this assumption affect the upper parts of the contours, i.e. the
upper limit of the amount of the lost material may be
overestimated by a few tenths $M_\odot$ if one assumes that the CE
contributes to the lost material. As Fig.~\ref{mlost} shows, the
mass-lose rate covers 3 orders of magnitude, i.e.\ from
$\sim10^{\rm -8}\,M_{\odot}\,{\rm yr}^{\rm -1}$ to $10^{\rm
-5}\,M_{\odot}\,{\rm yr}^{\rm -1}$. The lower limit of the
mass-loss rate for the CEW model is mainly constrained by the mass
loss during helium shell burning, while the upper limit is mainly
determined by the Eddington luminosity. The lower mass-loss rate
from systems in the CEW model may help to explain why the
circumstellar environment contains very little mass around some
SNe Ia such as SN 2011fe (\citealt{PATAT13}), although there are
ongoing arguments on whether SN 2011fe originates from a SD or a
DD system (\citealt{LIWD11c}; \citealt{CHOMIUK13};
\citealt{MAZZALI14}; \citealt{MENGXC16}).

In our model, a large amount of material may be lost from the CE
surface: it may be as large as $2.5\,M_{\odot}$. This could be the
cause of the colour excess seen in SNe Ia (\citealt{MENG09b}). As
this material cools, part of it may form dust and cause light
echoes (\citealt{WANGXF08}; \citealt{YANGY15}). The dust should be
distributed over a large region around SNe Ia, where the extent
should be smaller than the maximum distance that the lost material
can reach. Fig.~ \ref{mlpc} shows the contour of the total amount
of material lost from the system and the maximum distance that the
lost material may reach by the time when $M_{\rm
WD}=1.378\,M_{\odot}$. The maximum distance is simply $d=v_{\rm
w}\times t_{\rm d}$, where $v_{\rm w}$ is the wind
velocity\footnote{For the present CEW model, the exact value of
the CEW velocity is still quite uncertain, but a value larger than
that of an AGB star may be expected. For example, the wind
velocity for an AGB star with a core of $0.8~M_{\odot}$ and an
envelope of $0.1~M_{\odot}$ is very likely to be smaller than that
of a system with a CO WD of $0.8~M_{\odot}$ and a companion of
$\sim 2.0~M_{\odot}$ covered by a CE of $10^{\rm -3}~M_{\odot}$,
e.g. the system shown in Fig.~\ref{ce0822}, due to a higher escape
velocity from the surface of the CE. In addition, a value of
$\leq100~{\rm km~s^{\rm -1}}$ for the outflow from the SD systems
is also consistent with the dynamics of the forward shock and the
X-ray emission from the shocked supernova ejecta in some SNRs
(\citealt{BADENES07}). } of the material lost, here assumed to be
50\,${\rm km}$ ${\rm s}^{\rm -1}$, and $t_{\rm d}$ is the delay
time from the onset of mass transfer to the moment when $M_{\rm
WD}=1.378\,M_{\odot}$. There is a trend in Fig.~ \ref{mlpc} that
the maximum distance is closer if a larger amount of material is
lost. This is caused by the effect of the initial WDs and the
companions on the mass-transfer time scale: for a lower initial WD
mass and a lower companion mass, the mass-transfer timescale for
increasing the WD to 1.378 $M_{\odot}$ is longer, and the wind
material can reach a larger distance. At the same time, a less
massive companion means a lower mass-transfer rate and
consequently much lower mass-loss rate, which results in a smaller
amount of the lost material although the mass-transfer timescale
is longer. In particular, for some SNe Ia, the amount of lost
material is quite small (no more than 0.2\,$M_{\odot}$), but is
spread over a very large scale ($\sim 300$ pc). In these cases,
the environment around the SNe Ia should be very clean. In a
follow-up paper, we will discuss the observational consequences of
these properties in a detailed binary population synthesis study.

We must emphasize that the results from Fig.~\ref{final} to
Fig.~\ref{mlpc} are based on the assumption that the WD explodes
as a SN Ia when its mass reaches 1.378\,$M_{\odot}$.  However, the
explosion could be delayed if the WD is rotating rapidly (see the
discussion in section \ref{subs:5.2}), in which case the final
state of the binary system could be quite different from what is
presented here. For example, the range of the companion mass
should move to a lower value and the range of the final orbital
period may be increased, i.e.\ range from a value lower than
0.2\,d to one longer than 6.5\,d. Similarly, the range of the
radius, rotation velocity and orbital velocity are also increased,
i.e.\ the lower boundary moves to a lower value while the upper
boundary moves to a higher value. As a consequence, $M_{\rm lost}$
would be larger, while the final mass-loss rate from the binary
system would be lower, producing a cleaner environment around the
SN Ia.

\section{Binary population synthesis}\label{sect:4}
Adopting the results in section \ref{sect:3}, we have estimated
the expected supernova frequency from the WD+MS channel using the
rapid binary evolution code developed by \citet{HUR00, HUR02}.
Hereafter, we use {\sl primordial} to refer to the binaries before
the formation of the WD+MS systems and {\sl initial} for the WD+MS
systems (see also \citealt{MENG09} and \citealt{MENGYANG10}).

\subsection{Common-envelope evolution}\label{subs:4.1}
In the evolution of a binary, the primordial mass ratio (the ratio
of the mass of the primary to the mass of the secondary) is
crucial for the nature of the first mass-transfer phase. There is
a critical mass ratio, $q_{\rm c}$, which determines the
evolutionary direction of a binary system. If the primordial mass
ratio is larger than $q_{\rm c}$, the system will experience
dynamically unstable mass transfer and the system will enter into
a CE phase \footnote{Note that the CE here forms on a dynamical
timescale while the CE in our CEW model is maintained on a thermal
timescale.} (Paczy\'{n}ski 1976). The value of $q_{\rm c}$ is a
function of the evolutionary state of the primordial primary at
the onset of RLOF (\citealt{HW87}; \citealt{WEBBINK88};
\citealt{HAN02}; \citealt{POD02}; \citealt{CHE08}). In this paper,
we set $q_{\rm c}$ = 4.0 if the primary fills its Roche lobe on
the MS or in the HG. Such a value is supported by various detailed
binary evolution studies (\citealt{HAN00}; \citealt{CHE02,
CHE03}). If the primordial primary fills its Roche lobe on the
first giant branch (FGB) or AGB, we adopt

\begin{equation}
q_{\rm c}=\left[1.67-x+2\left(\frac{M_{\rm c1}^{\rm P}}{M_{\rm
1}^{\rm P}}\right)^{\rm 5}\right]/2.13,  \label{eq:qc}
  \end{equation}
where $M_{\rm c1}^{\rm P}$ is the core mass of the primordial
primary, and $x={\rm d}\ln R_{\rm 1}^{\rm P}/{\rm d}\ln M_{\rm
1}^{\rm p}$ is the mass-radius exponent of the primordial primary,
where $x\simeq0.3$ and changes with the stellar composition. If
the mass donors (primaries) are naked helium giants, we take
$q_{\rm c}$ = 0.748 (see details in \citealt{HUR02}).

The ``new'' binary embedded in the CE consists of the dense core
of the primordial primary and the primordial secondary and will
lose its orbital energy due to the friction between the CE and the
binary system. A part of the orbital energy released by the system
during the spiral-in process is injected into the envelope to
eject the material in the CE (\citealt{LS88}). In this paper, we
assume that the CE is ejected completely if
\begin{equation} \alpha_{\rm CE}\Delta E_{\rm orb}=|E_{\rm
bind}|,
  \end{equation}
where $E_{\rm bind}$ is the binding energy of the CE, $\Delta
E_{\rm orb}$ is the orbital energy released during the spiral-in
phase, and $\alpha_{\rm CE}$ denotes the CE ejection efficiency
(i.e.\ the fraction of the released orbital energy that can be
used in ejecting the CE). Since the internal energy and, in
particular, the recombination energy in the envelope is not
incorporated in the binding energy
(\citealt{IVANOVA13,IVANOVA15}), $\alpha_{\rm CE}$ can be larger
than 1 (see \citealt{HAN95} on the discussion of the internal
energy). Here, $\alpha_{\rm CE}=1.0$ or $\alpha_{\rm CE}=3.0$ is
adopted.

\subsection{Evolutionary channels for WD + MS systems}\label{subs:4.2}
There are three evolutionary channels to form WD + MS systems
based on the evolutionary state of the primordial primary at the
onset of the first RLOF.

Case 1 (He star channel): the primordial primary is in the HG or
on the RGB at the onset of the first RLOF phase (so-called case B
evolution as defined by \citealt{KW67}). In this case, a CE forms
either because of a large mass ratio or because the mass donor has
a deep convective envelope. After CE ejection, the system consists
of a helium star and a MS star. The helium star continues to
evolve and will fill its Roche lobe again after helium has been
exhausted in the center. Due to the small mass ratio, mass
transfer is dynamically stable leading to the formation a close CO
WD + MS system (see \citealt{NOM99, NOM03} for details).

Case 2 (EAGB channel): the primordial primary is on the early AGB
stage (EAGB) (i.e.\ helium is exhausted in the core, while thermal
pulses have not yet started). At this stage, the
hydrogen-exhausted core, i.e. the CO core surrounded by a dense
helium shell, is much denser than the hydrogen-rich envelope, and
there is a steep gradient in the gravitational potential between
the hydrogen-exhausted core and the hydrogen-rich envelope (e.g.
Figure. 2 in \citealt{MENG08}). If the system experiences
dynamically unstable mass transfer, the system will enter a CE and
spiral-in phase where the hydrogen-rich envelope forms a CE. After
the ejection of the CE, the new, much closer binary consists of
the hydrogen-exhausted core and the secondary. Eventually the
primordial primary becomes a helium red giant and may fill its
Roche lobe to start a second phase of RLOF. Similarly to the He
star channel, this RLOF phase is stable and produces WD + MS
systems after the end of RLOF.

Case 3 (TPAGB channel): the primordial primary fills its Roche
lobe during the TPAGB stage. Similarly to the above two channels,
a CE is formed during the first RLOF phase. A CO WD + MS binary is
produced after CE ejection.

The WD + MS systems continue to evolve, and the MS companions
will, at some stage, fill their Roche lobes and transfer matter to
the CO WDs, which may subsequently explode as SNe Ia. Here, we
assume that, if the initial orbital period, $P_{\rm orb}^{\rm i}$,
and the initial secondary mass, $M_{\rm 2}^{\rm i}$, of a WD + MS
system are located in the appropriate region in the ($\log P^{\rm
i}, M_{\rm 2}^{\rm i}$) plane (see Fig.~\ref{grid}) for SNe Ia at
the onset of RLOF, a SN Ia is produced.

\subsection{Basic parameters for the Monte-Carlo simulations}\label{subs:4.3}
To investigate the birth rate of SNe Ia, we followed the evolution
of $10^{\rm 7}$ binaries using the Hurley rapid binary evolution
code (\citealt{HUR00, HUR02}). The results of the grid
calculations in section \ref{sect:3} are incorporated into the
code. The primordial binary samples are generated in a Monte-Carlo
way, where circular orbits are assumed for all binaries. The basic
parameters for the simulations are as follows:

(i) The initial mass function (IFM) of \citet{MS79} is adopted.
The primordial primary is generated according to the formula of
\citet{EGG89}
\begin{equation}
M_{\rm 1}^{\rm p}=\frac{0.19X}{(1-X)^{\rm 0.75}+0.032(1-X)^{\rm
0.25}},
  \end{equation}
where $X$ is a random number in the range [0,1], and $M_{\rm
1}^{\rm p}$ is the mass of the primordial primary, which is
between 0.1\,$M_{\rm \odot}$ and 100\,$M_{\rm \odot}$.

(ii) For the evolution of a binary system, the primordial mass
ratio, $q'$, is a crucial parameter as it determines the
evolutionary direction of the system. In the paper, a uniform
mass-ratio distribution is adopted, i.e $q'$ is uniformly
distributed in the range  $[0,1]$ (\citealt{MAZ92};
\citealt{GM94}):
\begin{equation}
n(q')=1, \hspace{2.cm} 0<q'\leq1,
\end{equation}
where $q'=M_{\rm 2}^{\rm p}/M_{\rm 1}^{\rm p}$.

(iii) All stars are assumed to be members of binary systems. For
the separation  of the binary systems, a constant distribution in
$\log a$ is assumed for wide binaries, while $a$ falls off
smoothly for close binaries:
\begin{equation}
a\cdot n(a)=\left\{
 \begin{array}{lc}
 \alpha_{\rm sep}(a/a_{\rm 0})^{\rm m} & a\leq a_{\rm 0};\\
\alpha_{\rm sep}, & a_{\rm 0}<a<a_{\rm 1},\\
\end{array}\right.
\end{equation}
where $\alpha_{\rm sep}\approx0.070$, $a_{\rm 0}=10\,R_{\odot}$,
$a_{\rm 1}=5.75\times 10^{\rm 6}\,R_{\odot}=0.13\,{\rm pc}$ and
$m\approx1.2$. The separation distribution adopted here implies an
equal number of wide binary systems per logarithmic interval, and
gives approximately 50 percent of binary systems with an orbital
period $\leq100$ yr (see also \citealt{HAN95}).

(iv) We simply adopt either a single starburst (i.e.\ $10^{\rm
11}\, M_{\odot}$ in stars is generated at a single instant of
time) or a constant star-formation rate $S$ (SFR) for the last
15\,Gyr. Here the value of $S$ is set to be $5\,M_{\odot}\,{\rm
yr^{-1}}$, which is calibrated to maintain the production of one
binary with $M_{\rm 1}>0.8\,M_{\odot}$ in the Galaxy every year
(see \citealt{IT84}; \citealt{HAN95}; \citealt{HUR02};
\citealt{WK04}). The value of $5\,M_{\odot}\,{\rm yr^{-1}}$ can
successfully reproduce the birth rate of the core-collapse
supernova and the $^{26}$Al 1.809-MeV gamma-ray line in the galaxy
(\citealt{TIM97}).

\section{Binary population synthesis results}\label{sect:5}

\begin{figure}
    \includegraphics[angle=270,scale=.35]{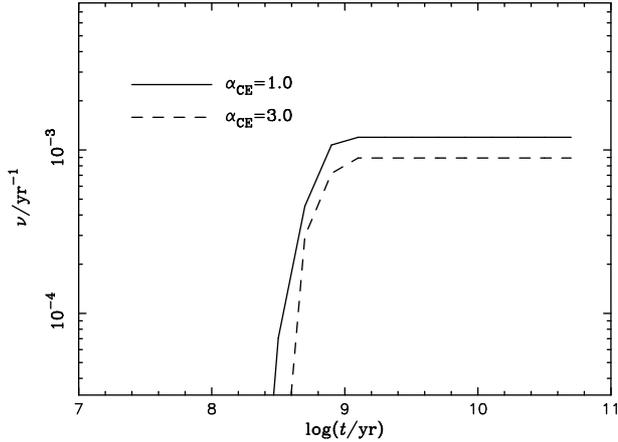}
\caption{The evolution of the birth rates of SNe Ia for a constant
star formation rate (Z=0.02, SFR=$5\,M_{\rm \odot}\,{\rm yr^{\rm
-1}}$). The solid and dashed curves show the cases with
$\alpha_{\rm CE}=1.0$ and $\alpha_{\rm CE}=3.0$, respectively.}
\label{sfr}
\end{figure}

\begin{figure}
    \includegraphics[angle=270,scale=.35]{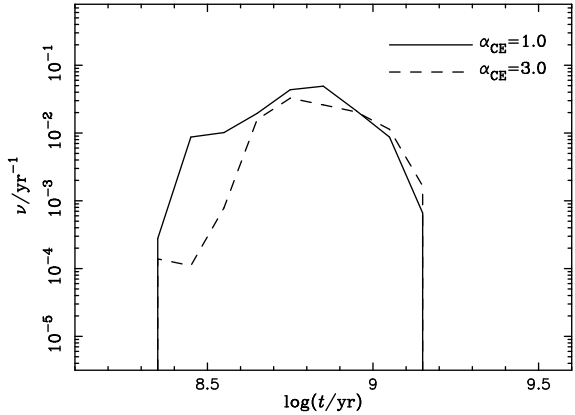}
\caption{The evolution of the birth rates of SNe Ia for a single
starburst of $10^{\rm 11}\,M_{\odot}$. The solid and dashed curves
show the cases with $\alpha_{\rm CE}=1.0$ and $\alpha_{\rm
CE}=3.0$, respectively.} \label{single}
\end{figure}
\subsection{The birth rate}\label{subs:5.1}
Fig.~\ref{sfr} shows the Galactic birth rates of SNe Ia (i.e.\
$Z=0.02$ and SFR= 5.0\,$M_{\odot}\,{\rm yr}^{\rm -1}$) from the
WD+MS channel.  The Galactic birth rate is around
(0.89-1.20)$\times10^{\rm -3}\,{\rm yr^{\rm -1}}$, which is higher
than that from the OTW model by about $\sim30\%$ (see Fig.~6 in
\citealt{MENG09}), as one would expect since the parameter space
for SNe Ia here is larger in our model than in the OTW model. This
result is somewhat lower but still comparable to that inferred
from observations (3-7$\times10^{\rm -3}\,{\rm yr^{\rm -1}}$,
\citealt{VAN91}; \citealt{CT97}; \citealt{LIWD11}).

The birth rate of SNe Ia for a single starburst is presented in
Fig.~\ref{single}. In this case, most supernovae occur between
0.25\,Gyr and 2\,Gyr after the starburst. Similarly, the peak
value here is also slightly higher than that from the OTW model
(see Fig.~7 in \citealt{MENG09}). In addition, Fig.~\ref{single}
shows that a low $\alpha_{\rm CE}$ leads to a higher birth rate
(see also Fig.~\ref{sfr}), since a low value of $\alpha_{\rm CE}$
means that a primordial system needs to release more orbital
energy to eject the CE for the formation of a WD + MS system;
hence the resulting WD + MS systems tend to have shorter orbital
periods and thereby more easily fulfill the conditions for SNe Ia.
In addition, only when $\alpha_{\rm CE}$ is low enough, may the WD
+ MS system produced from the TPAGB channel contribute to the SN
Ia population (see also \citealt{MENG09}). Moreover, our CEW model
may produce SNe with slightly shorter time delays compared to the
OTW model as the initial companion can be more massive.

\begin{figure}
    \includegraphics[angle=270,scale=.35]{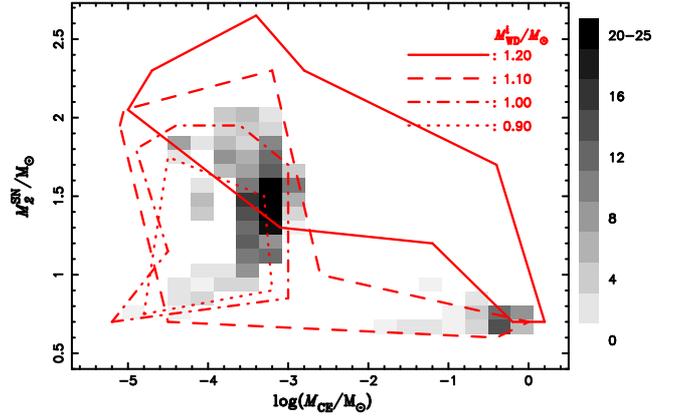}
\caption{The distribution of the CE mass and the companion mass
when $M_{\rm WD}=1.378\,M_{\odot}$, and the contour in the
($M_{\rm CE}$-$M_{\rm 2}^{\rm SN}$) plane for different initial WD
masses. Here, we only show the case of $\alpha_{\rm CE}=1.0$ since
the contours for the case of  $\alpha_{\rm CE}=3.0$ are very
similar.} \label{m2mce}
\end{figure}

\subsection{The CE mass at the time of the supernova explosion}\label{subs:5.2}
Fig.~\ref{grid} shows that some SNe Ia can explode in the CEW
phase, i.e.\ there is still some material in the CE when $M_{\rm
WD}=1.378\,M_{\odot}$, which could affect the observational
properties of SNe Ia. This is the most significant difference
between the CEW and the OTW model. Fig.~\ref{m2mce} shows the
distribution of the CE mass and the companion mass when $M_{\rm
  WD}=1.378\,M_{\odot}$ and the contour of the CE mass and the
companion mass in a $M_{\rm 2}^{\rm SN}-\log (M_{\rm CE})$ plane.
Here, we only show the case of $\alpha_{\rm CE}=1.0$ since the
case of $\alpha_{\rm CE}=3.0$ is similar. Roughly 1.2\%$-$14.2\%
of all SNe Ia explode in the CEW phase, which depends on the value
of $\alpha_{\rm
  CE}$. The percentage for which SNe Ia explode in the CEW phase in
the CEW model is slightly larger than the one for which SNe Ia
explode in the OTW phase in the OTW model (\citealt{MENG10};
\citealt{MENGYANG11b}). The CE mass ranges from $10^{\rm
  -5}\,M_{\odot}$ to $\sim 1\,M_{\odot}$. For most cases, the CE mass
lies between a few $10^{\rm -4}\,M_{\odot}$ and a few $10^{\rm
  -3}\,M_{\odot}$. It may be difficult to detect such hydrogen-rich
material directly; however, such low-mass CEs might leave
footprints in the high-velocity features in the spectrum of SNe Ia
(\citealt{MAZZALI05a,MAZZALI05b}).

In some cases, the CE mass can be larger than 0.1\,$M_{\odot}$,
even as large as 1\,$M_{\odot}$. If the SNe Ia occurs in such
situations, hydrogen lines should easily be detectable in the
early spectra of some SNe Ia. While such strong hydrogen lines
were indeed discovered in some SNe Ia, such as SN 2002ic
(\citealt{HAM03}), this is not consistent with observations of
most SNe Ia. This may require a time delay between the moment when
$M_{\rm WD}=1.378\,M_{\odot}$ and the supernova explosion; this
time needs to be long enough so that the CE material can be lost.
Based on the CE mass and the mass-reduction rate of the CE, we
estimate that at least a few $10^{\rm 5}$\,yr is needed for such a
delay time. Obviously, a simmering phase, lasting $\sim 10^{\rm
3}$\,yr, of the WD before the supernova explosion does not produce
such a delay time (\citealt{PIRO08}; \citealt{CHENXF14}). A
possible suggested mechanism is provided by the spin-up/spin-down
model (\citealt{JUSTHAM11}; \citealt{DISTEFANO12}), where the WD
is spun up by accretion. If the resulting WD's rotation velocity
is sufficiently large, the WD may not explode as a SN Ia even when
its mass exceeds 1.378\,$M_{\odot}$.  Hence to explode, a
spin-down phase of the WD is required. During the spin-down phase,
the companion may lose almost all of its hydrogen-rich envelope
and become a very faint object. Such a mechanism may explain the
lack of hydrogen lines in the late-time spectrum of SNe Ia
predicted by the interaction between the supernova ejecta and a
hydrogen-rich companion, and why no surviving companion has been
found, e.g., in the supernova remnant SNR 0509-67.5
(\citealt{JUSTHAM11}; \citealt{DISTEFANO12}). However, from a
purely theoretical point of view, it is completely unclear how
long the spin-down timescale is (\citealt{DISTEFANO11}). Recently,
based on the observational fact that CSM exists around some SNe
Ia, \citet{MENGPOD13} provided a constraint on the spin-down
timescale using a semi-empirical method; they found that the
spin-down timescale should be shorter than a few $10^{\rm 7}$\,yr,
otherwise it would be impossible to detect the signature of the
CSM. Such a timescale is long enough to erase most of the possible
observational signatures predicted by the SD model, but short
enough that it does not affect the delay time of supernova
explosions\footnote{The delay time is the time that has elapsed
between the formation of the primordial binary and the supernova
(SN) event.} since the formation of the primordial system
formation. $10^{\rm 5}$\,yr is needed for our CEW model to fulfill
the constraint in \citet{MENGPOD13}. Because of the possibility of
relatively massive initial companions (in the top right corner in
Fig.~\ref{grid}), the delay time for SNe Ia exploding in the CE
phase with $M_{\rm CE}>0.1\,M_{\odot}$ ranges from 240\,Myr to
330\,Myr. We estimate that no more than 1.7 in 100 SNe Ia belong
to this subclass. Those SNe Ia exploding in massive CEs may
contribute to 2002ic-like SNe Ia since their birth rate and delay
time is consistent with those estimated from observations, and a
strong hydrogen line is expected from the interaction between the
supernova ejecta and a massive CE (\citealt{ALD06}).

In Fig.~\ref{m2mce}, there is a gap in the distribution of $M_{\rm
CE}$, which is mainly caused by the different evolutionary stages
of the companion stars at the onset of RLOF which sets the
mass-transfer timescale. Generally, the mass-transfer timescale
for companion stars more massive than $\sim 2~M_{\odot }$ does not
vary monotonically with evolutionary stage from the zero-age main
sequence to the bottom of the RGB (\citealt{POD02}). Consequently,
the critical mass ratio for dynamically unstable mass transfer
also does not vary monotonically with the evolutionary stage,
which itself is determined by the orbital period of the system at
the beginning of RLOF (\citealt{GEHW15}).  Generally, for a given
binary system, a longer mass-transfer timescale means a lower
average mass-transfer rate.  If RLOF begins when the companion
crosses the HG, the exact evolutionary stage which gives the
longest mass-transfer timescale depends on the initial companion
mass. Systems with the longest mass-transfer timescale are more
likely to be found in the SSS phase when $M_{\rm
WD}=1.378~M_{\odot}$ rather than in the CEW phase, although
systems with the same initial WD and companion masses but with
shorter or longer orbital periods may explode in the CEW phase or
experience a delayed dynamical instability (see panel (7) in
Fig.~\ref{grid}).

For those with longer orbital periods, the mass-transfer timescale
sharply decreases with orbital period, which results in a very
high mass-transfer rate; then the CE mass may be larger than
$0.1~M_{\odot }$ at the time of the supernova explosion, an
example of which is shown in Fig.~\ref{110633}. In addition, for
very high mass-transfer rates, the companion will lose its outer
envelope quickly. With the loss of the companion's envelope, the
hydrogen fraction of the transferred material decreases, which
results in a higher critical accretion rate and hence a higher
mass-growth rate of the CO WD (see equation (\ref{eq:mdotcr}) and
panels (2) and (3) in Fig.~\ref{110633}). So, the time from the
onset of RLOF to the moment of the supernova explosion is
shortened; this also contributes to the population of SNe with
massive CEs as the timescale for losing the CE is also shortened.
However, such cases only occur for systems with initial WD masses
$\geq1.1~M_{\odot}$ and relatively massive companions.

For those with shorter initial orbital periods, the mass-transfer
timescale may become so short that the systems experience a
delayed dynamical instability if the mass ratio of the donor star
to the WD is sufficiently high. Such systems may eventually merge
and hence not contribute to the SN Ia population. If the mass
ratio is not very high, the mass-transfer timescale of the systems
with a shorter initial orbital period will become shorter, but not
much shorter than the longest mass-transfer timescale. This
implies a relatively higher, but not much higher mass-transfer
rate.  So, a massive CE is difficult to maintain for very high
mass-loss rates from the surface of the CE. Even for the cases
almost experiencing a delayed dynamical instability, the massive
CE can also not be maintained for a long time (see also
Fig.~\ref{delay}).  At the moment when $M_{\rm
WD}=1.378~M_{\odot}$, the mass-transfer rate of these systems is
generally slightly larger than the critical accretion rate, which
means that the CE cannot be very massive, e.g. $\leq 10^{\rm
-3}~M_{\odot }$. Additionally, the failure to maintain a massive
CE at the time of the supernova explosion is also partly caused by
the relatively longer mass-growth timescales of the WDs, even for
systems with the same initial WD mass, due to the unchanged
hydrogen mass fraction (see the different explosion times in
Figs.~\ref{110633} and \ref{delay}). SNe Ia that explode in a
low-mass CE predominantly occur for systems with initial WD masses
$\leq1.0~M_{\odot}$, where the mass-growth timescale of the WDs is
too long to maintain a very massive CE at the time of the
explosion. Therefore, if a system explodes in the CEW phase, the
CE tends to have a mass of either $\leq 10^{\rm -3}~M_{\odot}$ or
$\geq0.1~M_{\odot }$, which leads to the gap in the distribution
of $M_{\rm CE}$. Moreover, as clearly shown by the red and green
points at the bottom of panel (7) in Fig. \ref{fin}, the
companions for those WDs exploding in a high-mass CE are less
massive than the ones exploding in a low-mass CE.

\begin{figure*}
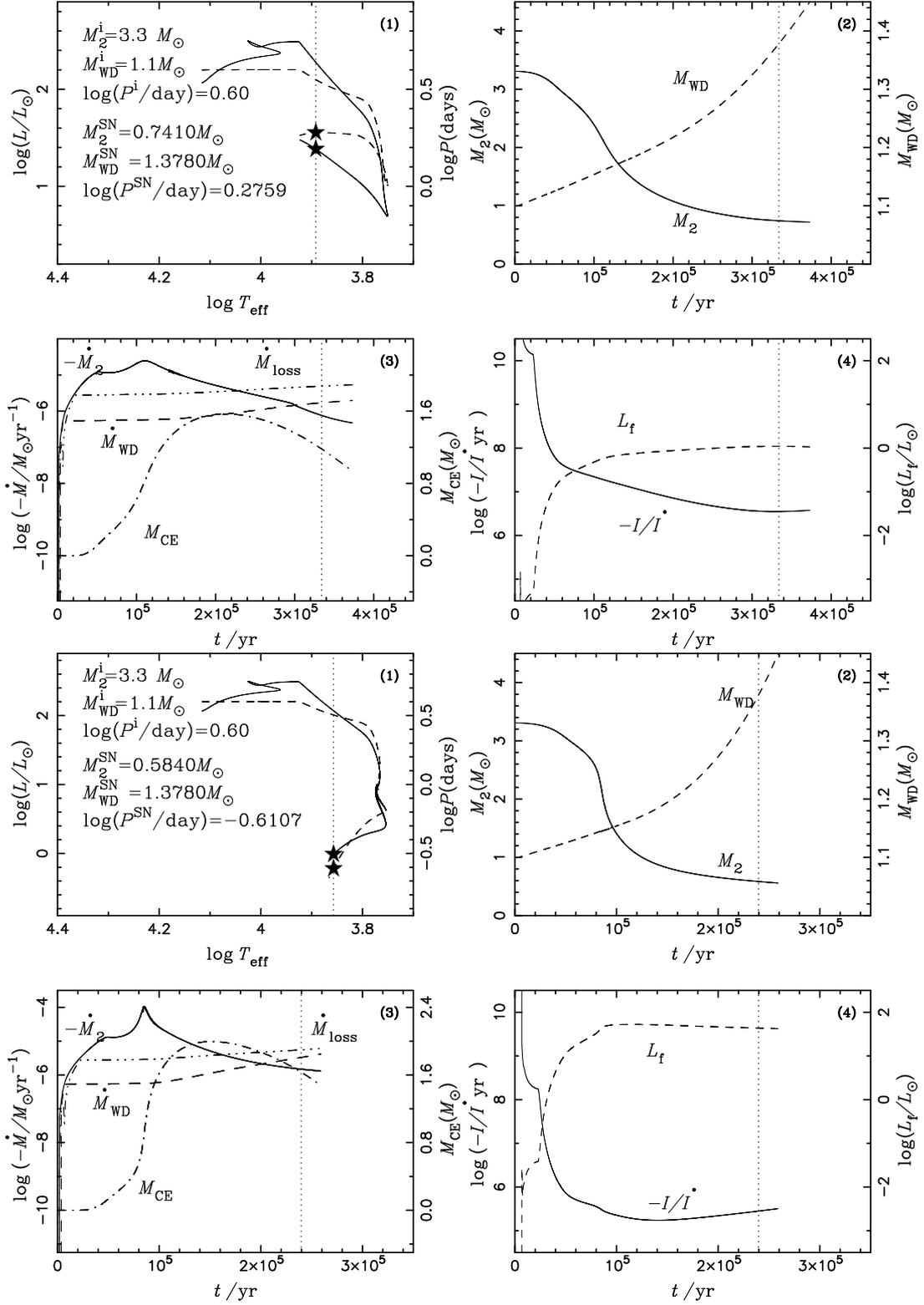

\vspace*{10mm}
    \includegraphics[angle=270,scale=.60]{den01.ps}
    \includegraphics[angle=270,scale=.60]{den10.ps}
\caption{Similar to Fig.~\ref{110633} but for a different average
CE density. The average density from equation (\ref{eq: rho}) is
arbitrarily multiplied by a factor of 0.1 (top four panels) and 10
(bottom four panels).} \label{den01}
\end{figure*}

\section{Discussion}\label{sect:6}
\subsection{Uncertainties in the CEW model}\label{subs:6.1}
In this paper, we constructed a new SD model in which a CE forms
around the binary when the  mass-transfer rate exceeds the
critical accretion rate. Material in the CE is lost from the
surface of the CE instead from the WD surface in the OTW model.
Many of the results in the CEW model are similar to those in the
OTW model, such as the evolution of binary parameters, while the
parameter space leading to SNe Ia and the birth rate of SNe Ia in
the CEW model are somewhat larger than in the OTW model. However,
the CEW model as presented in this paper is still very simple at
present with numerous uncertainties in the detailed modelling. In
the following, we discuss some of the main uncertainties in our
new CEW models one by one.

\begin{figure}
    \includegraphics[angle=270,scale=.35]{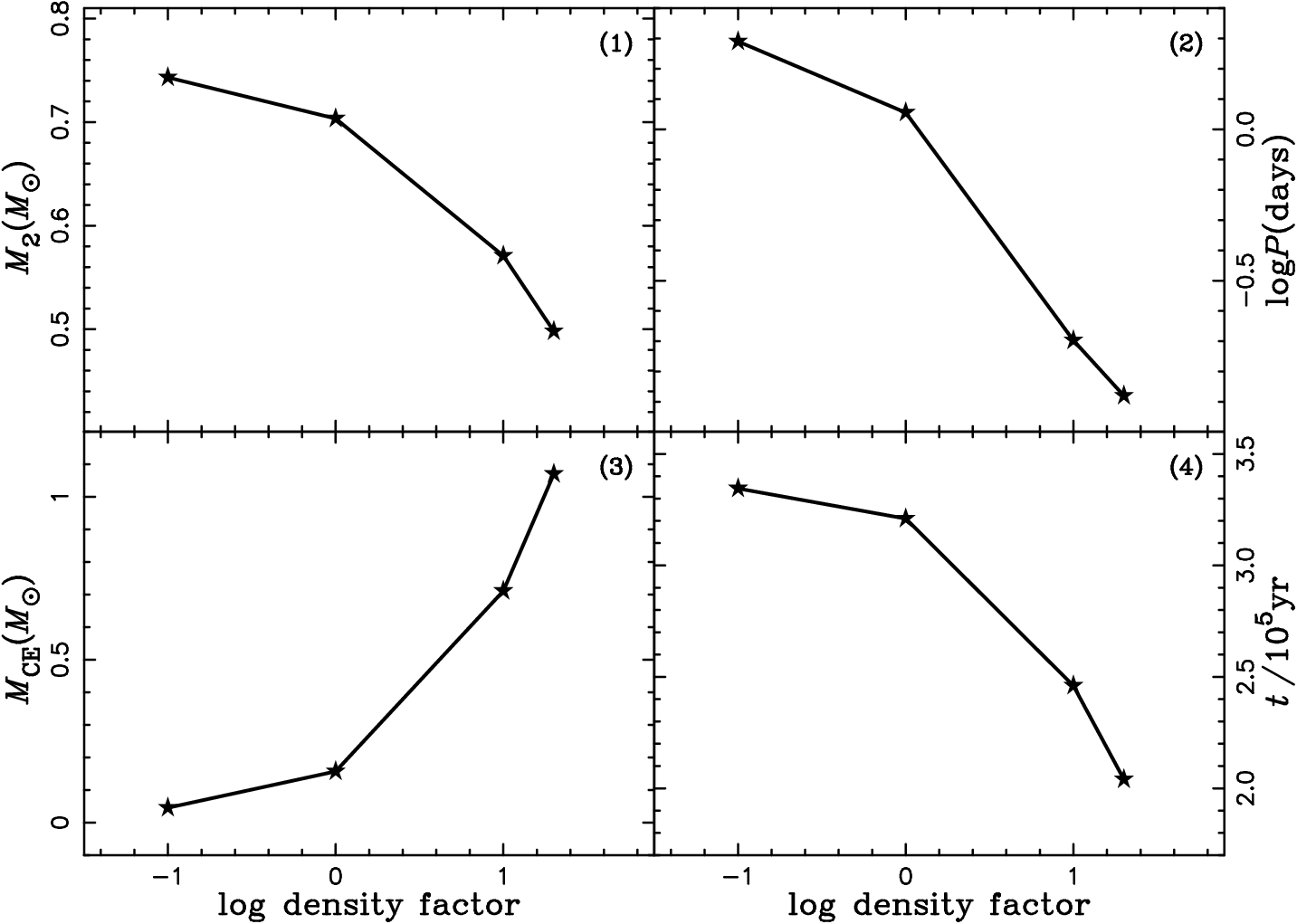}
\caption{The companion mass, orbital period, CE mass and
mass-transfer timescale (from the onset of RLOF to the supernova
explosion) when $M_{\rm WD}=1.378\,M_{\odot}$ as a function of the
density factor, where the initial parameter of the model is
($M_{\rm WD}^{\rm i}/M_{\odot}$, $M_{\rm 2}^{\rm i}/M_{\odot}$,
$\log(P^{\rm i}/{\rm d})$)=(1.1, 3.3, 0.6).} \label{denfac}
\end{figure}

In the CEW model, one of the main uncertainties arises from the
assumed CE density. In this paper, we simply assumed that the CE
is spherical and used an average density to replace the density at
the boundary of the inner and the outer part of the CE. The
density $\rho$ directly affects the effective turbulent viscosity
(equation \ref{eq:eta}), and hence the angular-momentum loss from
the binary system, the frictional luminosity and consequently the
mass loss from the system. To test the effect of varying the CE
density on the final parameter contours leading to SNe Ia, we
arbitrarily multiplied the density in equation (\ref{eq: rho}) by
factors of 0.1 and 10, respectively, and recalculated the contour
leading to SNe Ia for $M_{\rm WD}^{\rm i}=1.1\,M_{\odot}$.
Fig.~\ref{den01} provides an example to show how the average
density affects the binary evolution. The figures show that, as
one might expect, a higher average density results in a higher
frictional luminosity and a higher mass-transfer rate, and hence
leads to a more massive CE. If the average density is high enough,
i.e.\ the density is increased by a factor of 20 or larger, the
system will experience a dynamical instability and the system will
eventually merge. The critical density factor is system-dependent.
For example, for the system with ($M_{\rm WD}^{\rm i}/M_{\odot}$,
$M_{\rm 2}^{\rm i}/M_{\odot}$, $\log(P^{\rm i}/{\rm d})$) = (0.8,
2.2, 0.4), the critical density factor is 250 due to a much less
massive CE. In addition, the evolutionary track of the companion
in the HR diagram is significantly affected by the frictional
density, as is the period. Fig.~\ref{denfac} shows how the density
factor affects the final state of the binary system when $M_{\rm
WD}=1.378\,M_{\odot}$: the companion mass, orbital period and the
mass-transfer timescale decrease, while the CE mass increases with
the density factor. This is a direct consequence of the higher
frictional luminosity as this leads to a larger loss of orbital
angular momentum due to the friction between the binary system and
the CE; this leads to a higher mass-transfer rate and a more
massive CE. At the same time, the companion will lose more
material and becomes less massive, i.e.\ may be as small as
0.5\,$M_{\odot}$ at the time when $M_{\rm WD}= 1.378$ $M_{\odot}$.
Due to the relatively small radius for the shorter orbital period,
the companion also has a low luminosity, i.e.\ is dimmer than
$1\,L_{\odot}$. The relatively short mass-transfer timescale for a
large density factor here is a direct consequence of equation
(\ref{eq:mdotcr}). A high density factor results in a higher
mass-transfer rate, and the companion will lose its outer envelope
more quickly. With the loss of the companion's envelope, the
hydrogen fraction of the transferred material decreases; this
leads to a higher critical accretion rate and a higher mass-growth
rate of the CO WD (see panels 2 and 3 in Figs.~\ref{110633} and
\ref{den01}) and a shorter overall mass-transfer timescale. The
influence of high density on the companion could provide some
clues in searches for surviving companions in supernova remnants,
e.g. a dim MS star with a relatively low surface hydrogen
fraction.

However, we find that the contours leading to SNe Ia in the ($\log
P_{\rm i}, M_{\rm 2}^{\rm i}$) planes for different average
densities is more or less the same as that shown in
Fig.~\ref{cour}, irrespective of the density factor, except that
more SNe Ia explode in the CE phase for a higher assumed CE
density. As shown in Fig.~\ref{denfac}, the higher the frictional
density, the more massive the CE, making it more likely that the
supernova explodes in the CE phase. As far as the contours
resulting in SNe Ia is concerned, only the upper boundary, which
is determined by systems that experience a delayed dynamical
instability, would be significantly affected by the frictional
density. For a given system, the higher the frictional density,
the more likely the system experiences a delayed dynamical
instability, moving the upper boundary downwards. However, as
discussed above, although this is model-dependent, only when the
density factor is larger than $\simeq20$ would systems that avoid
the merging fate in our standard model experience a delayed
dynamical instability. Therefore, within the density range tested
here, the upper boundaries of the contours are almost unaffected
by the density factor. The left and right boundaries are
determined by the radius of stars on the ZAMS, or at the the end
of the HG, which are completely unaffected by the density factor.
The lower boundary is constrained by the condition that the
secondaries are massive enough to increase the WD mass to
$1.378~M_{\odot}$. This condition is very weakly dependent on the
frictional density factor as discussed above for a system with
($M_{\rm WD}^{\rm i}/M_{\odot}$, $M_{\rm 2}^{\rm i}/M_{\odot}$,
$\log(P^{\rm i}/{\rm d})$) = (0.8, 2.2, 0.4).

\begin{figure*}
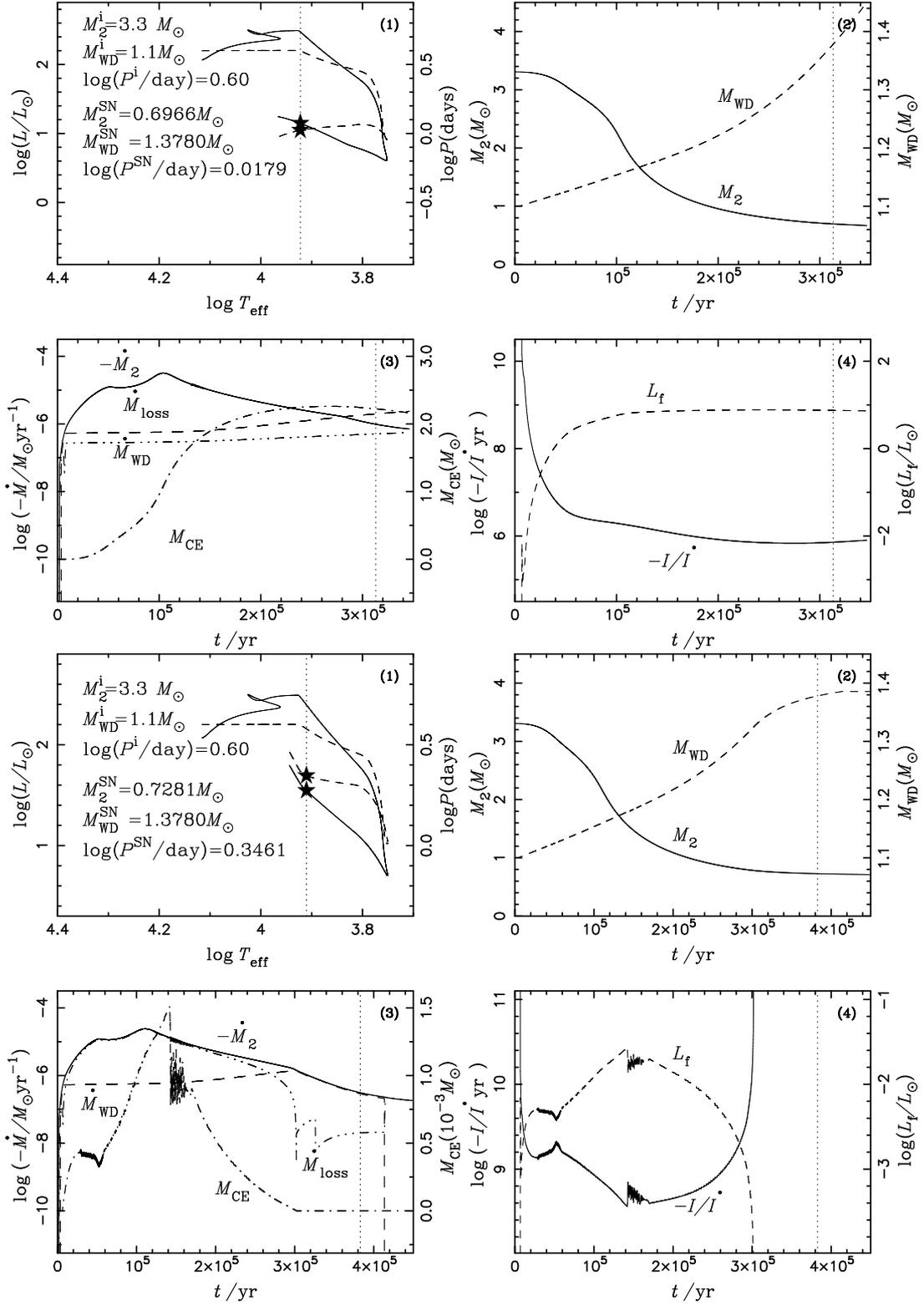

\vspace*{0mm}
    \includegraphics[angle=270,scale=.60]{wind01.ps}
    \includegraphics[angle=270,scale=.60]{wind10.ps}
\caption{Similar to Fig.~\ref{110633} but for a different
mass-loss rate. The mass-loss rate from equation (\ref{eq:wind1})
is arbitrarily multiplied by a factor of 0.1 (top four panels) or
10 (bottom four panels).} \label{wind01}
\end{figure*}

\begin{figure}
    \includegraphics[angle=270,scale=.35]{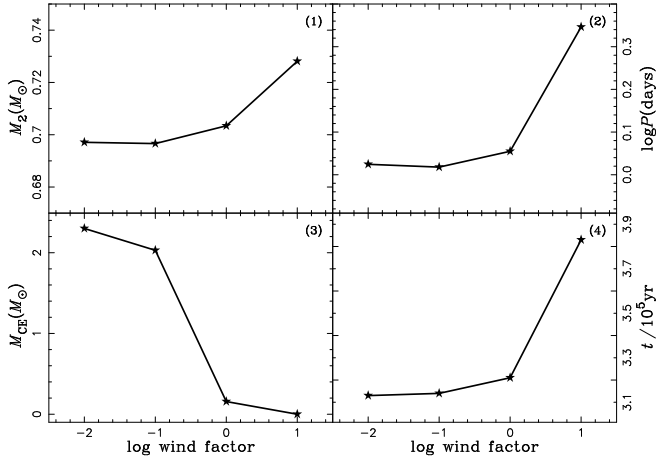}
\caption{The companion mass, orbital period, CE mass and
mass-transfer timescale (from the onset of RLOF to the supernova
explosion) when $M_{\rm WD}=1.378\,M_{\odot}$ as a function of the
wind factor, where the initial parameter of the model is ($M_{\rm
WD}^{\rm i}/M_{\odot}$, $M_{\rm 2}^{\rm i}/M_{\odot}$,
$\log(P^{\rm i}/{\rm d})$)=(1.1, 3.3, 0.6).} \label{winfac}
\end{figure}

Another key uncertainty arises from the adopted mass-loss rate
from the surface of the CE, which determines the mass left in the
CE; there is no well established prescription for mass loss from a
CE (cf.~\citealt{SCHRODER07}). Here, we again arbitrarily multiply
the mass-loss rate in equation (\ref{eq:wind1}) by a factor of 0.1
or 10. Fig.~\ref{wind01} shows an example of how the mass-loss
rate affects the binary evolution and the final state of the
binary system. Figs.~\ref{110633} and \ref{wind01} show that the
wind mass-loss rate does not affect the binary evolution as
significantly as the density does, while it may influence the CE
mass considerably. There is a clear trend that, for a lower wind
mass-loss rate, WDs tend to reach 1.378\,$M_{\odot}$ during the CE
phase, while WDs are more likely to explode in the SSS or RN phase
for the higher mass-loss rates. In other words, for a higher
mass-loss rate, it becomes more difficult to maintain the CE while
it is relative easy to obtain a more massive CE for a lower
mass-loss rate. In fact, when the wind enhancement factor is
larger than 10, the effect of the CE wind on the evolution of the
binary is rather minor, and the evolution is very similar to that
of the OTW. However, the WD may gradually increase its mass to
1.378\,$M_{\odot}$ for any mass-loss rate. Fig.~\ref{winfac} shows
how the wind enhancement factor affects the final state of the
binary system. The wind mass-loss rate only has a minor effect on
the SN formation rate when the wind mass-loss rate is sufficiently
low (the wind factor is lower than 0.1), although a lower wind
factor favours the existence of the CE.  This directly follows
from equation (\ref{eq:mcedot}), i.e.\ $\dot{M}_{\rm
CE}=|\dot{M}_{\rm 2}|-\dot{M}_{\rm WD}-\dot{M}_{\rm
wind}\simeq|\dot{M}_{\rm 2}|-\dot{M}_{\rm WD}$ when the wind
mass-loss rate is much lower than $\dot{M}_{\rm 2}$ and
$\dot{M}_{\rm WD}$ and can therefore be neglected. Thus, if the
wind factor is low enough, the evolution of the CE no longer
depends on the mass-loss rate.

Fig.~\ref{denwin} presents a map for SNe Ia and mergers in the
density -- wind factor plane; the boundary between SNe Ia and
mergers is also shown in the figure. The figure illustrates that
the boundary is system-dependent, especially for the density
factor. However, for a rather large parameter region, binary
systems may avoid the fate of merging, demonstrating that the CEW
model is rather robust.

The factor $\alpha$ in Equation (1) is also an important factor
which directly affects the orbital angular-momentum loss and the
frictional luminosity. Its influence on the binary evolution is
quite similar to the changes in the frictional density as already
discussed above.

In this paper, we assumed that all the potential energy sources,
i.e.\ from nuclear burning, the secondary's luminosity and the
frictional luminosity are contributing to driving the wind mass
loss. A small part of that energy will be required to drive the
expansion of the CE itself, which could moderately decrease the
expected mass loss.  However, as we discussed above, there is a
rather large parameter range for which the assumptions about the
wind do not affect whether a binary system explodes as a SN Ia or
not. Therefore, this effect is unlikely to change the main
conclusions of this paper.

We also assumed that the presence of an envelope changes the
helium flash at the bottom of the hydrogen-burning shell like in a
TPAGB star. On the other hand, if the CE is not massive enough and
the helium flash is strong enough, the CE might be destroyed. We
plan to study the conditions for which the destruction of the CE
can be prevented in the future.

Finally, for the CEW model it is still unclear what the system
actually would look like during the CE phase; this will depend on
the CE mass, the nuclear luminosity for hydrogen and helium
burning and the percentage of the total energy used to drive the
mass loss from the surface of the CE. In general, we would expect
a RG-like object with a high luminosity. We will address this
issue in the future following the approach developed previously in
\citet{POD90}.

In summary, our CEW model as it stands now is still very simple,
but most of our assumptions are rather conservative. We therefore
consider the results in this paper only as the first step and we
intend to refine the model in the future.

\subsection{The differences between the CEW model and the OTW model}\label{subs:6.2}
The physical assumptions in the CEW model and the OTW model are
quite different, which should lead to several testable differences
between the two models.

1) The CEW model does not directly depend on metallicity; hence it
allows SNe Ia to occur in any environment irrespective of
metallicity and redshift. Thus, the discovery of some SNe Ia at
high redshifts and/or in a low-metallicity environment are easily
explained in the CEW model. The lack of a low-metallicity
threshold for SNe Ia relative to SNe II is also a natural result
in the CEW model.

2) When $|\dot{M}_{\rm 2}|$ exceeds $\dot{M}_{\rm cr}$, the
hydrogen-rich material burns stably on the WD at a rate of
$\dot{M}_{\rm cr}$, while the unprocessed material forms the CE.
At the same time, some CE material may leave the system from the
surface of the CE. This treatment leads to two significant
differences between the CEW model and the OTW model. One is that
the mass-growth rate of the WD in the CEW model is larger than
that in the OTW model, which results in a larger initial parameter
space for SNe Ia and a lower minimum WD mass, and hence a higher
birth rate. The other is the wind velocity. In the CEW model, the
unprocessed material leaves the system from the surface of the CE
instead of the surface of the WD, resulting in a much lower wind
velocity than in the OTW model, i.e.\ most likely less than 200
${\rm km}\,{\rm s}^{\rm -1}$, probably lower than 50 ${\rm
km}\,{\rm s}^{\rm -1}$. Our model is therefore compatible with the
dynamics of the forward shock and the X-ray emission from the
shocked ejecta in SN Ia remnants examined by \citet{BADENES07}.
Even the properties of RCW 86, where the wind cavity with a
diameter of $\sim 30$ pc contains 1.6\,$M_{\odot}$ of material
ejected from the progenitor system (\citealt{BROERSEN14}), may be
explained by our model since a low-density cavity around some SNe
Ia could also be expected for some cases in our model grid (see
Fig.~\ref{mlpc}).

3) When the hydrogen-rich material is stably burning on the
surface of the WD, soft X-ray emission is expected from the
mass-accreting WD and the system will appear as a supersoft source
(SSS, \citealt{VANDERHEUVEL92}), while no strong X-ray emission is
expected from the DD scenario until just before the SN Ia
explosion. This provides a potential method for distinguishing
between SD and DD models. However, the number of the SSSs in
elliptical and spiral galaxies is much smaller than that predicted
from the standard SD model (\citealt{GB10};
\citealt{DISTEFANO10}). The predicted number of SSSs from the SD
model in these studies is based on the assumption that all the
accreting WDs are in the SSS phase which lasts for about two
million years before the SN Ia explosion. This assumption is
rather simple and may overestimate the duration of the SSS phase
by a factor of 10 (see panel 3 in Figs~\ref{ce0822} and
\ref{110633}). According to the OTW model, in which the OTW may
absorb the soft X-rays, the duration of the SSS phase is much
shorter; \citet{HKN10} and \citet{MENGYANG11b} were able to
reproduce the observed SSS number in elliptical and spiral
galaxies, i.e.\ only a small fraction of mass-accreting WDs
resulting in SNe Ia contribute to the supersoft X-ray flux of a
galaxy. In the CEW model, because of the existence of the CE,
X-rays would be completely absorbed by the CE or the material lost
from the CE surface. The duration of the SSS phase in the CEW
model would be even shorter than in the OTW model since, even when
$|\dot{M}_{\rm 2}|$ is smaller than $\dot{M}_{\rm cr}$, a CE can
still exist (see the panel 3 in Fig.~\ref{110633}). So, the CEW
model is likely to be consistent with the observations of
\citet{GB10} and \citet{DISTEFANO10} (also see the discussion
about the SSS phase in \citealt{MAOZ14}).

4) The OTW model may explain the properties of some RNe which have
been suggested as potential progenitors of SNe Ia (\citealt{HK05,
  HK06a, HK06b}; \citealt{HKL07}); this also applies to the CEW model
as the CEW model reduces to the OTW model when there is no CE.

\begin{figure}
    \includegraphics[angle=270,scale=.35]{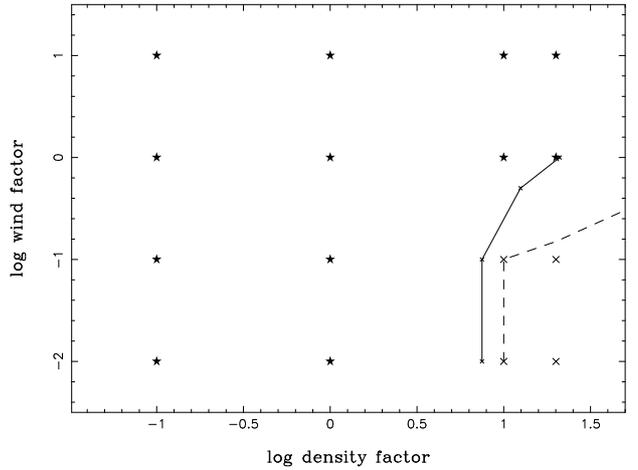}
\caption{The map for SNe Ia and mergers in the density -- wind
factor plane (for a representative system with initial parameters
of ($M_{\rm WD}^{\rm i}/M_{\odot}$, $M_{\rm 2}^{\rm i}/M_{\odot}$,
$\log(P^{\rm i}/{\rm d})$)=(1.1, 3.3, 0.6) and (0.8, 2.2, 0.4),
respectively). Stars represent systems that explode as SNe Ia,
while crosses indicate systems that merge. The solid and dashed
curves show the boundary between SNe Ia and mergers obtained from
the systems of (1.1, 3.3, 0.6) and (0.8, 2.2, 0.4), respectively.}
\label{denwin}
\end{figure}

\begin{figure}
    \includegraphics[angle=270,scale=.35]{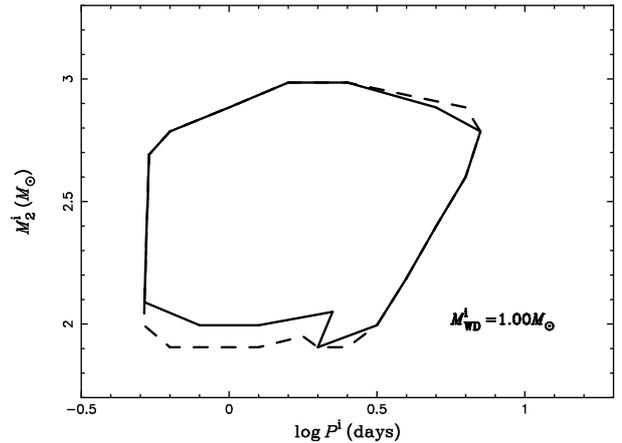}
\caption{Contours leading to SNe Ia in the secondary mass --
orbital period plane for $M_{\rm WD}^{\rm i}=1.0\,M_{\odot}$. The
solid curve is for $\eta_{\rm He}$ in \citet{KH2004} and the
dashed curve is for $\eta_{\rm He}$ arbitrarily increased by
50\,\%, i.e.\ $\eta_{\rm He}^{\rm '}=\min(1, 1.5\eta_{\rm He})$.}
\label{comhe}
\end{figure}

5) The efficiency of the accretion of hydrogen on massive WDs at a
high rate over many cycles is still controversial; this affects
the conditions for which SNe Ia can occur. For example,
\citet{IDAN12} recently concluded that the accumulated helium
under the hydrogen-burning shell is completely lost in a strong
helium flash, which, taken at face value, would make SNe Ia
impossible. On the other hand, \citet{NEWSHAM13} conclude that a
WD may continue to increase its mass toward the Chandrasekha
limit.
%
The helium accumulation efficiency, $\eta_{\rm He}$, used here is
usually smaller than 1 even when hydrogen burns at a rate of
$\dot{M}_{\rm cr}$. Fortunately, although it is not clear how the
WD adjusts to alternating helium flashes and stable hydrogen shell
burning in the OTW model, this issue probably does not arise in
the CEW model. In addition, $\eta_{\rm He}$ could be larger than
derived by \citet{KH2004} if the effect of H shell burning has
been accounted for, since the nuclear energy produced via
H-burning keeps the underlying He buffer hotter, resulting in
weaker and more frequent He flashes (\citealt{CASSISI98}).
Especially, \citet{HILLMAN15} recently found that, although a
significant fraction of the material accumulated prior to the
flash is lost during the first few helium shell flashes, the
fraction decreases with repeated helium shell flashes and
eventually no mass is lost at all during subsequent flashes. In
addition, the helium shell flash in a rapidly rotating WD due to
accretion will become more stable than in the corresponding
non-rotating WD (\citealt{YOON04a,YOON04b}). We tested the effect
of the accumulation efficiency of helium flashes on the parameter
space leading to SNe Ia by increasing $\eta_{\rm He}$ by 50\%
rather arbitrarily (but keeping the maximum value of $\eta_{\rm
He}$ at 1). Fig.~\ref{comhe} shows a comparison of the contours
leading to SNe Ia for different values of $\eta_{\rm He}$. The
figure shows that a higher $\eta_{\rm He}$ moves the lower
boundary to a lower value while the other boundaries are almost
unaffected by a change of $\eta_{\rm He}$. The reason is that
$\eta_{\rm He}$ only plays a role in phases of stable burning or
weak hydrogen flashes (see Fig.~\ref{grid}). In addition, for a
lower initial WD mass ($<0.9\, M_{\odot}$), most SNe Ia explode in
the SSS or RN phase, where $\eta_{\rm He}$ may play a more
important role, i.e.\ a high $\eta_{\rm He}$ results in a larger
parameter space for SNe Ia.  So, a higher $\eta_{\rm He}$ could
lead to a relatively higher birth rate of SNe Ia and produce SNe
Ia in an older population.

6) As far as the possible appearance of a WD in the OTW phase is
concerned, \citet{NOM07} suggested that the system might look like
an undersized OB star, i.e.\ with an effective temperature of
$31600 - 10000$ K, but with a size of only a few $R_{\odot}$,
limited by the WD's RL radius. However, no such rapidly accreting
blue white dwarf has been found in searches in the Large
Magellanic Cloud (\citealt{LEPO13}). Alternatively such systems
might look like Wolf-Rayet (WR) stars or WD planetary nebulae (see
also \citealt{LEPO13}; \citealt{NIELSEN13}; \citealt{WHELLER13}).
It is worth emphasizing that, for most cases, when the
mass-transfer rate is very high, i.e.\ higher than $10^{\rm
-5}\,M_{\odot}$\,yr$^{\rm -1}$ (see \citealt{HAN04} and
\citealt{MENG09}), the OTW must carry away a large amount of
dynamic energy which almost equals the nuclear burning energy. So,
during the OTW phase, the WD must be dimmer. In the CEW model, we
suggest that the system resembles a TPAGB star during the CE
phase. If the system loses the envelope before the supernova, it
could look like a post-AGB star.
%
%
After the CE phase, the accreted hydrogen-rich material burns
stably on the surface of the WD, and the system may show the
properties of a SSS, which is a powerful source of ionizing
ultraviolet emission. Such ultraviolet photons may ionize helium
in the CSM or interstellar medium, and then emission in the
recombination lines of HeII could be expected. Interestingly,
searching for the recombination radiation of HeII in passively
evolving galaxies, \citet{JOHANSSON14} found that the strength of
the observed HeII $\lambda$4686 nebular emission is consistent
with the post-AGB stars being the sole ionizing source. In
addition, during the CE phase, if the envelope is massive enough,
a high luminosity,
%
%
i.e.\ even higher than the Eddington luminosity of the WD, and a
variable high mass-loss rate for helium flashes might be expected.
LMC N66 in the Large Magellanic Cloud could be an interesting
candidate: it consists of a WD primary and a non-degenerate system
in the center and experiences huge mass loss with a highly
variable mass-loss rate (\citealt{LEPO13b}). In any case, it is
still unclear how to distinguish these from real TPAGB stars, and
a detailed study on the appearance of the binary system for our
CEW model is still required. We have started such an investigation
and will report the results in a future publication.

7) One question that has been raised for a long time since
\citet{HAC96,HAC99b} is whether the OTW model requires too much
fine-tuning (e.g. \citealt{PIERSANTI00}; \citealt{SHEN07};
\citealt{WOOSLEY11}). In contrast, as discussed in section
\ref{subs:6.1}, our CEW model is rather robust and may work within
a very large parameter range, probably requiring much less
fine-tuning (see Fig.~\ref{denwin}).

\section{Summary and conclusion}\label{sect:7}
We have constructed a new SD model for SNe Ia in which a common
envelope forms when the mass-transfer rate onto the WD exceeds a
critical accretion rate instead of an optically thick wind. Based
on the CEW model, we investigated in detail the WD+MS channel
using detailed binary evolution calculations. The initial
parameters for SNe Ia in the ($\log P^{\rm i}, M_{\rm 2}^{\rm i}$)
plane were obtained and then used to study the evolution of the
birth rate of SNe Ia by a BPS method. In the following we
summarize the main results.

1) The parameter space obtained from our new CEW model is slightly
larger than that in the OTW model, i.e.\ it allows a higher
initial companion mass and longer initial orbital period. The
minimum WD mass leading to SNe Ia is 0.65\,$M_{\odot}$, slightly
lower than in the OTW model.

2) We also presented the final parameters of the binary systems
and the companion properties at the time of the supernova
explosion, which may be helpful for identifying progenitor systems
or search for surviving companions in supernova remnants. Several
observed SSSs and RN systems lie in the parameter regions for SNe
Ia, making them ideal candidates for SN Ia progenitors.  The
properties of Tycho G, the suggested surviving companion of
Tycho's supernova, is also consistent with our results.

3) Before the SN Ia explosion, the system may lose as much as
2.5\,$M_{\odot}$ of hydrogen-rich material, where there is a trend
that, the more mass is lost, the closer the maximum distance this
material may reach. For some SNe Ia, the material may be less than
0.2\,$M_{\odot}$, spread over a very large region, possibly
extending to more than 300 pc. The environment around these SNe Ia
could be very clean and it would be very difficult to detect a
signature of this CSM around such SNe Ia.

4) A significant fraction of SNe Ia may explode in the CE phase
with a CE mass from a few $10^{\rm -4}\,M_{\odot}$ to a few
$10^{\rm
  -3}\,M_{\odot}$; these might be essential for explaining the
high-velocity Ca features seen in the spectrum of SNe Ia
(\citealt{MAZZALI05a,MAZZALI05b}).

5) For a single star burst, most SNe Ia occur between 0.25\,Gyr
and 2\,Gyr, while the Galactic birth rate of the SNe Ia from our
CEW model is (0.89-1.20)$\times10^{\rm -3}\,{\rm yr^{\rm -1}}$,
which is higher than in the OTW model by about $\sim 30\,$\%.

 6) We examined the uncertainties of our model caused by uncertainties
 in the modeling parameters and found that the model is very robust
 within a very large parameter range.

7) We discussed the differences between the CEW and the OTW model.
Our model avoids several of the shortcomings of the OTW model but
also shares the merits of the OTW model. Especially, our model
does not depend on metallicity, and SNe Ia may occur in
low-metallicity and/or high-redshift environments. In addition,
the typical outflow velocity should be much lower than in the OTW
model since the material is lost from the CE surface instead of
the WD surface in the OTW model.

8) The accumulation efficiency for helium flashes could be a very
important factor in determining the parameter space that leads to
a successful SN Ia,  i.e.\  a higher $\eta_{\rm He}$ than in
\citet{KH2004} may extend the allowed range of companion masses to
a lower initial mass, resulting in a higher birth rate and a
possibly longer delay time.

9) Our model may naturally explain the birth rate and delay time
of 2002ic-like SNe Ia.

\section*{Acknowledgments}
We are grateful to the anonymous referee for his/her constructive
comments which helped to improve the manuscript greatly. We thank
Professor Zhanwen Han for his helpful comments on the manuscript.
This work was partly supported by Natural Science Foundation of
China (Grant Nos. 11473063, 11522327), CAS (No. KJZD-EW-M06-01),
CAS ``Light of West China'' Program and Key Laboratory for the
Structure and Evolution of Celestial Objects, Chinese Academy of
Sciences.







\newpage
\appendix
\section[]{The method for calculating the CE mass in our code}
In section \ref{subs:2.2}, equations (\ref{eq:mcedot}),
(\ref{eq:wind1}), (\ref{eq:rg}) and (\ref{eq:cemass}) form a
coupled set of equations that determine the mass in the CE. Here,
we show how this is determined in our code. The main idea is that
the formation of the CE and the mass loss from the CE surface may
be divided into two steps, i.e. the CE forms first, and then the
wind begins to blow. In our code, there is a transition value
$M_{\rm CE}^{'}$, which is determined by
 \begin{equation}
M_{\rm CE,i+1}^{'}=M_{\rm CE, i}+(|\dot{M}_{\rm 2}|_{\rm
i}-\dot{M}_{\rm WD,i})\cdot\Delta t,\label{eq:cemtr}
  \end{equation}
where the initial value of $M_{\rm CE, 0}=0$. Then, putting the
transition value into Equation (\ref{eq:rg}), we may get the CE
radius. Based on equations (\ref{eq:eta}), (\ref{eq:lf}) and
(\ref{eq: rho}), the frictional luminosity $L_{\rm f}$ may be
obtained, and then the total luminosity $L_{\rm tot}=L_{\rm
nuc}+L_{\rm 2}+L_{\rm f}$, where $L_{\rm nuc}$ is nuclear energy
for stable hydrogen burning and $L_{\rm 2}$ is the secondary
luminosity. $L_{\rm nuc}$ is obtained from equation
(\ref{eq:Lnu}). Therefore, according to Equation (\ref{eq:wind1}),
we may get the mass-loss rate $\dot{M}_{\rm wind}$, where the CE
mass is also the transition value from Equation (\ref{eq:cemtr}).
The CE mass is then calculated according to
 \begin{equation}
M_{\rm CE,i+1}=M_{\rm CE,i+1}^{'}-\dot{M}_{\rm wind,i}\cdot\Delta
t.\label{eq:cem}
  \end{equation}
Equation (\ref{eq:cemtr}) + Equation (\ref{eq:cem}) deduces to
 \begin{equation}
 \displaystyle
  \begin{array}{lc}
M_{\rm CE,i+1}=M_{\rm CE, i}+(|\dot{M}_{\rm 2}|_{\rm
i}-\dot{M}_{\rm WD,i}-\dot{M}_{\rm wind,i})\cdot\Delta
t\\
\\
 \hspace{1.25cm} =M_{\rm CE, i}+\dot{M}_{\rm CE,i}\cdot\Delta t,\label{eq:cemf}
 \end{array}
  \end{equation}
which is exactly the same as Equation (\ref{eq:cemass}).


\bsp    
\label{lastpage}
\end{document}